\documentclass[12pt]{article}

\usepackage{amsmath}
\usepackage{amsfonts}
\usepackage{amssymb}
\newtheorem{thm}{Theorem}
\newtheorem{cor}[thm]{Corollary}
\newtheorem{lemma}[thm]{Lemma}

\newtheorem{prop}[thm]{Proposition}

\newtheorem{rem}[thm]{Remark}

\usepackage{graphicx}
\usepackage{bm,bbm}

\usepackage{epsfig}
\usepackage{latexsym}
\usepackage{url}
\usepackage{bm,bbm}
\def\noi{\noindent}

\def\tr{\hbox{tr}}
\def\calm{{\mathcal{M}}}

\def\calh{{\mathcal{H}}}

\def\calc{{\mathcal{C}}}

\def\calv{{\mathcal{V}}}

\def\bra{\langle}
\def\ket{\rangle}
\def\raw{\rightarrow}

\def\pmx{\begin{pmatrix}}
\def\emx{\end{pmatrix}}

\def\R{\mathbb{R}}
\def\C{\mathbb{C}}
\def\dg{\dagger}
\def\co{{\rm conv}}
\def\Tr{{\rm Tr}}
\def\vol{{\rm vol}}
\def\vrad{{\rm vrad}}

\date{}

\begin{document}

\title
{Geometry of sets of quantum maps: a generic positive map acting
  on a high-dimensional system is not completely positive}

\author
{Stanis{\l}aw J. Szarek$^{1,2}$, Elisabeth Werner$^{1,3}$,
and  Karol \.Zyczkowski$^{4,5}$
\smallskip \\
$^1${\small Case Western Reserve University, Cleveland, Ohio, USA} \\
$^2${\small Universit{\'e} Paris VI, Paris, France} \\
$^3${\small Universit{\'e} de  Lille 1, Lille, France} \\
$^4${\small Institute of Physics, Jagiellonian University, Krak{\'o}w,
Poland}\\
$^5${\small Center for Theoretical Physics, Polish Academy of Sciences,
Warsaw, Poland}}

\date{December 21, 2007}

\maketitle

\begin{abstract}
We investigate the set a) of positive, trace preserving maps
acting on density matrices of size $N$,
and a sequence of its nested subsets:
the sets of maps which are b) decomposable, c) completely positive,
d) extended by identity impose positive partial transpose and e) are
superpositive.
Working with the Hilbert-Schmidt (Euclidean) measure
we derive tight explicit two-sided bounds for the volumes 
of all five sets. A sample consequence is the fact that, as  $N$ increases,
a generic positive map becomes not decomposable and, {\em a fortiori},
not completely positive.
Due to the Jamio{\l}kowski isomorphism, the results obtained for quantum 
maps are closely connected to similar relations between the volume of the 
set of quantum states and the volumes of its subsets (such as 
states with positive partial transpose or separable states) or supersets. 
Our approach depends on
systematic use of duality to derive {\em quantitative} estimates,
and on various tools of classical 
convexity, high-dimensional  probability and geometry of Banach 
spaces, some of which are not standard.
\end{abstract}

\newpage

\section{Introduction}

Processing of quantum information takes place in physical laboratories,
but it may be conveniently described in a finite dimensional Hilbert space.
The standard set of tools of a quantum mechanician
includes density operators which represent physical states. 
A density operator $\rho$ is Hermitian, positive semi-definite and normalized.
The set of density operators of ``size" $2$ is equivalent,
with respect to the Hilbert-Schmidt (Euclidean) geometry, 
to a three ball, usually called the Bloch ball. The set 
of density operators of ``size"  $N$
forms an ${N^2-1}$-dimensional  convex body which naturally
embeds into ${\calm}_N$, the space of $N \times N$ 
(complex) matrices.

The interesting geometry of these non-trivial, high--dimensional
sets attracts a lot of recent attention \cite{ACH93, HHH96a,
AMM97,BH01,EMMM01}.
In particular one computed their Euclidean volume 
and hyper-area of their surface  \cite{ZS03},
and investigated properties of its boundary \cite{GMK05}.

If the dimension $N$ of the Hilbert space ${\mathcal H}_N$ is a composite
number, 
the density operator can describe a state of a bipartite system.
If such a state has the tensor product structure,
$\rho=\rho_A\otimes \rho_B$, then it represents uncorrelated subsystems. 
In general, following \cite{werner}, a state is called {\sl separable} if it
can 
be written as a convex combination of product states.
In the opposite case the state is called {\sl entangled}
and it is valuable for quantum information processing \cite{NC00},
since it may display non--classical correlations.

The set  ${\calm}_N^{\rm sep}$ of separable states
forms a convex subset of positive volume of the entire set of states,  
which we will denote by  ${\calm}_N^{\rm tot}$ \cite{ZHSL98}.
Some estimations of the relative size 
of the set of separable states were obtained in 
\cite{BCJLPS99,Zy99,GB,Sza04,Sl04,AS,Hi07}, 
while its geometry was analyzed in 
\cite{LS98,STV98,KZ01,SBZ06}. Similar issues for 
infinite-dimensional systems were studied in \cite{CH00}.

Quantum information processing is inevitably
related with dynamical changes of the physical system.
Transformations that are discrete in time can be described
by linear {\sl quantum maps}, or {\em super-operators}, $\Phi: {\calm}_N \to {\calm}_N$ 
(or, more generally, $\Phi: {\calm}_K \to {\calm}_N$).
A map is called {\sl positive} (or {\em positivity-preserving})
if any positive (semi-definite) operator
is mapped into a positive operator. 
A map $\Phi$ called {\sl completely positive}  (CP)
if the extended  map $\Phi \otimes \ \mathcal {I}_k$
is positive for any size $k$ of the extension. Here $\mathcal{I}_k$ is the identity map on 
${\mathcal M}_k$. 
%(The symbol ${I}_k$ will stand fr the identity on $\mathbb{C}^k$, 
%i.e., for the  $ k \times k$ identity matrix.)
We will denote the {\em cones} of positive and completely positive maps
(on ${\calm}_N$) by $\mathcal{P}_N$ and $\mathcal{CP}_N$ 
respectively, or simply  by $\mathcal{P}$ and $\mathcal{CP}$ 
if the size of the system is fixed or clear from the context.

Conservation of probability in physical processes
imposes the {\sl trace preserving} (TP) property:
Tr\,$\Phi(\rho)={\rm Tr}\,\rho$. 
It is a widely accepted paradigm that any physical process 
may be described by a {\sl quantum operation}: 
a completely positive, trace preserving map. 
(In the context of quantum communication, quantum operations 
are usually called {\em quantum channels}.)

The set ${\cal CP}_N^{\rm TP}$ of quantum operations,
 which act  on density operators of size $N$,
forms a convex set of dimension $N^4-N^2$.
Due to Jamio{\l}kowski isomorphism \cite{Ja72,ZB04}
the set $N^{-1}{\cal CP}_N^{\rm TP}$ can be considered as a subset
of the $(N^4-1)$--dimensional set ${\calm}_{N^2}^{\rm tot}$
of density operators acting on an extended 
Hilbert space, ${\cal H}_N \otimes {\cal H}_N$.
This useful fact contributes to our understanding
of properties the set of quantum operations,
but its geometry is nontrivial even
in the simplest case of $N=2$ \cite{RSW02,BZ06}.

The main aim of the present work is to 
derive tight two-sided bounds for the Hilbert--Schmidt (Euclidean)
volume of the set  ${\cal CP}_N^{\rm TP}$ of quantum operations
acting on density operators of size $N$ 
and analogous estimates for the
volume of the sets ${\cal P}_N^{\rm TP}$ of positive trace preserving maps, 
and of similar subsets of the {\em superpositive} cone ${\cal SP}_N$ 
(see (\ref{dualityA}) and/or \cite{An04}) or
the cone ${\cal D}_N$ of {\em decomposable} maps (see (\ref{decompos}))  etc.
We show that, for large $N$,
some subsets cover only a very small fraction of its immediate superset, 
while in some  other cases the gap between  volumes is relatively small.
These bounds are related to (and indeed derived from, making use 
of the Jamio{\l}kowski  isomorphism) analogous  relations between 
the volumes of various subsets of the set of quantum states
such as those consisting of separable states or of states with {\em positive
partial transpose}  (PPT) (see the paragraph following  (\ref{coneT}))
and their dual objects.
Our methods are quite general and allow to produce 
tight two-sided estimates for many other sets of quantum 
states or of quantum maps.

\smallskip
The paper is organized as follows. In the next section
we introduce some necessary definitions 
involving the set of trace preserving positive maps 
and its relevant subsets or supersets, which will allow us to 
present an overview of the results obtained in this paper
(summarized in Tables 2-4).
Section \ref{prelim} contains more definitions  and various 
preliminary results. Most of those results are not new, but 
many of them are not well-known in
the quantum information theory community. 
In section \ref{results} we state precise versions of our results
and outline their proofs.
Some details of the proofs and technical results (from all sections) are 
relegated to Appendices.

\newpage
\section{Positive and trace preserving maps: notation and overview of results}

\subsection{Cones of maps and matrices}

Let $\Phi: \mathcal{M}_N \raw \mathcal{M}_N$ be a linear quantum map,
or a super-operator. 
More general maps $\Phi: \mathcal{M}_K \raw \mathcal{M}_N$ 
may also be considered and analyzed by essentially the same 
methods, but we choose to focus on the case $K=N$ to limit 
proliferation of parameters. 

\smallskip Let $ \rho \in \mathcal{M}_N$;   
the transformation $\rho'= \Phi(\rho)$ can be  described by 
\begin{equation}
\rho'_{n \nu}\ = \  
 \Phi_{\stackrel{\scriptstyle n\nu}{m \mu} }
  \, \rho_{m \mu}  \, ,
 \label{dynmatr1}
\end{equation}
where we use the usual Einstein summation convention.
 The pair of upper indices  $^{n \nu}$ defines
its ``row," while the lower indices  $_{m \mu}$
determine the ``column." This agrees with the usual linear algebra
convention of representing linear maps as matrices. The  
relevant basis of  $\mathcal{M}_N$ is here 
$E_{ij} := |e_i\ket \bra e_j|$, $i,j=1,\ldots,N$, where $(e_i)_{i=1}^N$ is an
orthonormal
basis of $\mathcal{H}_N$ (which can be identified with $\C ^N$), and the
$m\mu$'th ``column" 
of $  \Phi_{\stackrel{\scriptstyle n \nu}{\scriptstyle m \mu} }$,  i.e., the $N \times N$
matrix
$ \big(\Phi_{\stackrel{\scriptstyle n \nu}{\scriptstyle m \mu}}\big)_{n, \nu = 1}^N$, is
indeed 
$\Phi(E_{m\mu}) = \Phi_{\stackrel{\scriptstyle n\nu}{\scriptstyle m \mu}} E_{n \nu}$\,.

\smallskip
By appropriately reshuffling elements of $  \Phi_{\stackrel{\scriptstyle n
\nu}{m \mu} }$
we obtain another matricial representation of a quantum map, the  dynamical
matrix 
$D_\Phi$  \cite{SMR61}, sometimes also called in the literature 
``the Choi matrix" of $ \Phi$. 
The  dynamical matrix is obtained  as follows
\begin{equation}
D_{\stackrel{ \scriptstyle m n }{\mu \nu}}
: =  \Phi_{\stackrel{\scriptstyle n \nu}{m \mu} }\,.
 \label{reshuff}
\end{equation}
An alternative (and useful) description of the dynamical matrix is as follows
$$
D_\Phi := \big( {\mathcal I}_N \otimes \Phi  \big)  \rho_{\rm max} 
= \sum_{m,\mu = 1}^N E_{m\mu} \otimes \Phi(E_{m\mu}),
$$
where $ \rho_{\rm max}  = |\xi \ket \bra \xi |$, with $|\xi \ket = \sum_{m=1}^N
e_m \otimes e_m$,
is a maximally entangled pure state on $\mathcal{H}_N \otimes \mathcal{H}_N$.

We point out that the order of indices of the matrix $D$ in (\ref{reshuff}) 
is different than in the previous work \cite{ZB04,BZ06}.
(The reason for this change will be elucidated in the next paragraph.)
  Note that in the present notation the operation of
 ``reshuffling," which converts matrix $\Phi$ into $D$,
 corresponds to a ``cyclic shift" of the four indices.

\smallskip
It is sometimes convenient to arrange the row and column indices of $D_\Phi$ 
($^{m n}$ and  $_{\mu \nu}$ respectively) in the lexicographic order, thus
obtaining a 
standard ``flat" $N^2 \times N^2$ 
matrix with a natural block structure: the leading indices
$_{\stackrel{\scriptstyle m}{ \mu}}$ 
indicate the position of the block and the second pair of indices 
$_{\stackrel{\scriptstyle n}{ \nu}}$ refers to the position of the entry within
a block.
In other words, the $m\mu$'th block of $D_\Phi$ is $\Phi(E_{m\mu})$ or
\begin{equation} \label{Choi}
D_\Phi = (\Phi(E_{m\mu}))_{m,\mu=1}^N ,
\end{equation}
an $N \times N$ block matrix with each block belonging to $\mathcal{M}_N$.

If a super-operator  $\Phi$ belongs to the positive cone $\mathcal{P}$  (i.e.,
$\Phi$ is 
positivity-preserving), 
then it also maps Hermitian matrices to Hermitian matrices. This in turn
is equivalent to $\Phi$ commuting with complex conjugation $^\dg$; 
in what follows we will generally consider only maps with this property.
It is easy to check that  Hermiticity-preserving is equivalent to 
the following relation 
(which has no obvious interpretation)
\begin{equation}
\Phi_{\stackrel{\scriptstyle n\nu }{m\mu }} =
\overline{\Phi_{\stackrel{\scriptstyle \nu n}{\mu m }}}.
 \label{symmetr}
\end{equation}
However, expressing condition (\ref{symmetr}) in terms of the dynamical matrix
we obtain
$$ 
D_{\stackrel{ \scriptstyle m n }{\mu \nu}}
 =  \overline{D_{\stackrel{ \scriptstyle \mu \nu }{m n}}} \ ,
 \label{symmetrdyn}
$$
which just means that $D_{\Phi}$ is Hermitian.
Thus one may describe linear  Hermiticity-preserving maps on $\mathcal{M}_N$
via Hermitian dynamical $N^2 \times N^2$ matrices. The property of being
positive 
can be characterized just as elegantly. 
A theorem of  Jamio{\l}kowski \cite{Ja72}
states that a map $\Phi$ is positive, $\Phi \in \mathcal{P}$,
if and only if the corresponding
dynamical matrix $D_\Phi$ is {\sl block positive}.
[A (square) block matrix $(M_{ij})$  (say, with $M_{ij} \in
\mathcal{M}_N$ for all $i,j$) is said to be block positive iff, for every
sequence of complex scalars $\xi=(\xi_j)$, the $N\times N$ matrix
$\sum_{i,j}M_{ij}\bar{\xi}_i \xi_j $ is positive semi-definite.]

\smallskip 
Arguably the most useful upshot of the dynamical matrix point of view 
arises in the study of CP maps.
A theorem of  Choi \cite{Cho75a} states that
a map $\Phi$ is completely positive, $\Phi \in \mathcal{CP}$, iff $D_\Phi$ is
positive
semi-definite. 
Therefore, to each CP map on $\mathcal{M}_N$ corresponds an $N^2 \times N^2$
(positive semi-definite) matrix,  and {\em vice versa}.
In particular, the rescaled dynamical matrix $D$ associated with a (non-zero)
CP map
represents a state of a bi--partite system,
$\sigma:=D/\Tr\, D \in {\calm}_{N^2}^{\rm tot}$ -- see e.g. \cite{Ja72,ZB04}, an
element 
of the {\em base } of the positive semi-definite cone obtained by intersecting 
that cone with the hyperplane of trace one matrices.

\smallskip 
If the dimension of the cones or other sets  
under consideration is relevant, 
we will explicitly use a lower index, writing, e.g., 
${\cal CP}_2$ for the set of one--qubit completely positive maps.

\subsection{Trace preserving maps}
The trace preserving property, Tr\,$\Phi(\rho)={\rm Tr}\rho$,
is equivalent to a condition for the {\em partial  trace} of the
dynamical matrix
 \begin{equation}
\sum_ n D_{\stackrel{\scriptstyle m n}{\mu n}} = \delta_{m\mu} \, ,
\hspace{5mm} {\rm or } \hspace{5mm} 
{\rm Tr}_B D = {I}_A \, .
\label{tpcond}
\end{equation}
Therefore the compact set  ${\cal CP}_N^{\rm TP}$ of quantum operations
may be defined as a common part of  the affine plane 
representing the condition  (\ref{tpcond})
and  the cone of positive semi-definite dynamical matrices - see Figure 1 
in section \ref{prelim}. 

In (\ref{tpcond}) and (occasionally) in what follows we use the labels $A, B$
to distinguish between the space on which the original state $\rho$ acts, 
namely $\calh_A$,  and  the space of $\Phi(\rho)$, denoted $\calh_B$.
In particular, $I_A$ stands for the identity operator on $\calh_A$.
Since such conventions are somewhat arbitrary 
(as was the ordering of indices of $D$), some care needs to be exercised when 
comparing (\ref{tpcond}) and similar formulae with other texts 
(such as, e.g., \cite{BZ06}).

\subsection{ Bases of cones} \label{bases}\vskip 3mm 
\noi
Let $H^0=\{M \in \mathcal{M}_d: \Tr \,M=0\}$. 
Next, let $H^{\rm b}=\{M \in \mathcal{M}_d: \Tr\, M=d^{1/2}\}$ and let 
$H^+=\{M \in \mathcal{M}_d : \Tr \,M\geq 0\}$. 
If $\calc \subset \mathcal{M}_d$ is a cone, we will denote by
$\mathcal{C}^{\rm b} := \mathcal{C} \cap H^{\rm b}$ the corresponding 
{\em base} of  $\mathcal{C}$.
(This definition makes good sense if $\calc \subset H^+$ or, equivalently,
if the $d \times d$  identity matrix $ {I}_d$ belongs to the dual cone $\calc^*$ 
(see (\ref{dualconeofmaps})). In this case
the 
cones generated by $\mathcal{C}^{\rm b}$ and $\calc$ coincide, perhaps 
after passing to closures.) 
%In fact in all the cases we will consider  the two cones are closed.)
We will use the same notation for the sets of quantum maps corresponding 
to matrices via the Choi-Jamio{\l}kowski isomorphism. Thus, for example,
$\Phi : \mathcal{M}_N \raw \mathcal{M}_N$ belongs to $H^{\rm b}$ iff 
$\Tr\, D_\Phi = \Tr \,\Phi({I}_N)=N$.
(Here the identity matrix $ {I}_N$ and its image $\Phi({I
}_N)$ are $N \times N$ matrices, while $D_\Phi$ is a $d \times d$ matrix,
with $d=N^2$; in particular the two trace operations take place in different
dimensions.)
Then $\mathcal{P} \cap H^{\rm b}=\mathcal{P}^{\rm b}$ is a base of the cone
$\mathcal{P}$, 
$\mathcal{CP}  \cap H^{\rm b}=\mathcal{CP}^{\rm b}$ is a base of the cone
$\mathcal{CP}$, 
and similarly for other cones that will be introduced later.  
The (real) dimension of the bases is $N^4-1$.

The reason behind our somewhat non-standard normalization $\Tr \,M=d^{1/2}$
is twofold. First, the condition can be rewritten as $\bra M,e\ket_{HS} = 1$,
where 
$e = {I}_d/d^{1/2}$ is a matrix whose Hilbert-Schmidt norm is equal to
one; this 
allows to treat  $e$ as a distinguished element of cones and -- at the same
time -- of their duals.  
Next, the primary objects of our analysis are quantum {\em maps}, 
and the chosen normalization assures that TP 
(and, dually, unital; see Appendix \ref{appE}) maps are in
$H^{\rm b}$.
When we are primarily interested in {\em states}, the normalization $\Tr \,M=1$
can be 
thought of as  more natural  (the distinguished element 
${I}_d/d$ is the then the maximally mixed state, usually denoted by
$\rho_*$).  
%[However, even in that context duality considerations 
%are somewhat awkward.]

While all the matrix spaces or spaces of maps are {\em a priori} complex, all
cones of interest will
 live in fact in the real space $\mathcal{M}_d^{\rm sa}$ of Hermitian matrices
or in the space of Hermicity-preserving maps. We will use the same symbols
$H^0$, $H^{\rm b}$ etc.  to denote the smaller {\em real } (vector or affine)
subspaces; this should not lead to misunderstanding.

\subsection{Other cones, all sets of interest compiled in one table}
 Analogous point of view will be employed when studying other cones 
of quantum maps such as

\smallskip \noi $\bullet$ the cone ${\mathcal SP}$ of {\em superpositive} maps
(also called {\em entanglement breaking}, 
see (\ref{dualityA}) and the paragraphs that follow)

\smallskip \noi $\bullet$ the cone ${\mathcal D}$ of {\em decomposable} maps 
(see (\ref{decompos}))

\smallskip \noi $\bullet$ the cone ${\mathcal T}$ of maps which extended by
identity 
impose positive partial transpose (see (\ref{coneT}) and the paragraphs that follow).  

\medskip \noi 
In all cases we will identify the corresponding cone of $N^2 \times N^2$
matrices 
and will relate in various ways bases of the cones and their sections 
corresponding to the trace preserving restriction. For easy reference, we list
all 
objects of interest in the table below;  see also Figures 1 and 3
in section \ref{prelim}. 
The missing definitions and unexplained 
relations (generally appealing to duality) will also be clarified there.

\begin{table}[ht]
\caption{Sets of quantum maps and the sets of quantum states associated to them
via 
the Jamio{\l}kowski--Choi isomorphism, cf. (\ref{isom0})-(\ref{isom3}) below. 
The inclusion relation holds in each collumn, e.g. 
${\cal P}_N \supset {\cal D}_N \supset {\cal CP}_N \supset {\cal T}_N \supset
{\cal SP}_N $.
The symbols $\circ$ and $\star$ in the rightmost column denote sets consisting
of also 
 non-positive semi-definite matrices  which technically are not states 
(and are not readily identifiable with objects appearing in the literature).
}
 \bigskip

{\renewcommand{\arraystretch}{1.67}
\begin{tabular}
[c]{|l| c  | c|}\hline \hline
{\quad  \quad Maps \hskip 0.67cm} & 
{\hskip 1.69cm} \  $\Phi:{\cal M}_N \to {\cal M}_N$  {\hskip 2.52cm}  &
States  $\sigma \in {\cal M}_{N^2}$   \\
\hline
\end{tabular}
\\
\begin{tabular}
[c]{|l| c c c c c | c|} 
 {} & cones & & ${\rm Tr}D_{\Phi}=N$ & &  ${\rm Tr}_B D_{\Phi}={I}_N$
& 
$\quad \quad {\rm Tr}\sigma \ = \ 1$ \quad   {\hskip -0.13cm} 
 \\ \hline 
\quad \ 
positive & ${\cal P}_N$ & $\supset$ & ${\cal P}_N^{\rm b}$ & $\supset$ & ${\cal
P}_N^{\rm TP}$ &
    $\circ$ \\
decomposable & ${\cal D}_N$ & $\supset$ & ${\cal D}_N^{\rm b}$ & $\supset$ &
${\cal D}_N^{\rm TP}$ & 
    $\star$ \\
\parbox{2.6cm}{\centering completely \\ \vskip-1mm positive}
  & ${\cal CP}_N$ & $\supset$ & ${\cal CP}_N^{\rm b}$ & $\supset$ & ${\cal
CP}_N^{\rm TP}$ & 
   ${\cal M}_{N^2}^{\rm tot}$ \\
\parbox{2.6cm}{\centering \vskip2mm PPT \\ \vskip-1mm inducing}
  & ${\cal T}_N$ & $\supset$ & ${\cal T}_N^{\rm b}$ & $\supset$ & ${\cal
T}_N^{\rm TP}$ & 
    ${\cal M}_{N^2}^{\rm PPT}$ \\ 
\parbox{2.6cm}{\centering \vskip2mm super \\ \vskip-1mm positive}
  & ${\cal SP}_N$ & $\supset$ & ${\cal SP}_N^{\rm b}$ & $\supset$ & ${\cal
SP}_N^{\rm TP}$ & 
   ${\cal M}_{N^2}^{\rm sep}$ \\
 \hline \hline
\end{tabular}
}
\label{tab1}
\end{table} 
%%%%%%%%%

\noi The action of the Jamio{\l}kowski--Choi isomorphism, associating cones of
maps  to cones of matrices and their respective bases, can be summarized as
\begin{eqnarray} 
 \Phi \hbox{  is positive (} \Phi \in {\cal P}_N )& \Leftrightarrow  &\sigma
\hbox{  is block-positive }
\label{isom0} 
\end{eqnarray}
 \begin{eqnarray} 
 \Phi \hbox{  is completely positive  (} \Phi \in {\cal CP}_N )&
\Leftrightarrow  &\sigma \hbox{  is positive semi-definite}
 \nonumber \\
\Phi \in {\cal CP}^{\rm b}_N \ & \Leftrightarrow & \ \sigma=\frac{1}{N}
D_{\Phi} \in {\cal M}^{\rm tot}_{N^2}
\label{isom1} 
\end{eqnarray}
 \begin{eqnarray} 
  \Phi \hbox{  is PPT inducing (} \Phi \in {\cal T}_N ) & \Leftrightarrow 
&\sigma \in \mathcal{PPT}
 \nonumber \\
\Phi \in {\cal T}^{\rm b}_N \ &\Leftrightarrow &\ \sigma=\frac{1}{N} D_{\Phi}
\in {\cal M}^{\rm PPT}_{N^2}
\label{isom2} 
\end{eqnarray} 
\begin{eqnarray} 
 \Phi \hbox{  is superpositive (} \Phi \in {\cal SP}_N ) & \Leftrightarrow 
&\sigma \hbox{  is separable }
 \nonumber \\
\Phi \in {\cal SP}^{\rm b}_N \ & \Leftrightarrow & \ \sigma=\frac{1}{N}
D_{\Phi} \in {\cal M}^{\rm sep}_{N^2}
\label{isom3} 
\end{eqnarray}
The description of the matricial cone  associated to the cone ${\cal D}_N$ of
decomposable maps
is largely tautological:  the sum of the positive semi-definite cone and its
image via the 
partial transpose.  We likewise have $\frac{1}{N} D_{\Phi} \in \circ$ (resp.,
$\in \star$) iff
$\Phi \in  {\cal P}^{\rm b}_N$ (resp.,  $\in {\cal D}^{\rm b}_N$).

\subsection{Comparing sets via volume radii, overview of results}

Explicit formulae for volumes of high dimensional sets  are often not very
transparent (when they can be figured out at all, that is). 
This may be exemplified by the closed expression for the volume of
the $d^2-1$--dimensional set ${\calm}_d^{\rm tot}$, the set of of density operators
of size $d$
that has been computed in \cite{ZS03} 
\begin{equation} 
\vol\big({\calm}_d^{\rm tot} \big) = \sqrt{d}\,(2\pi)^{d(d-1)/2}
\frac{\Gamma(1)\dots\Gamma(d)}{\Gamma(d^2)} .
\label{explicit}
\end{equation}

%\smallskip
Given the complexity of formulae such as (\ref{explicit}),
the following concept is sometimes convenient. Given an $m$-dimensional set
$K$,
we define ${\rm vrad}(K)$, the  {\em volume radius} of $K$, as the radius of an 
Euclidean ball of the same volume (and dimension) as $K$. Equivalently, 
${\rm vrad}(K)= \big(\vol(K)/\vol (B_2^m)\big)^{1/m}$, where $B_2^m$ 
is the unit Euclidean ball. It is fairly easy (if tedious) to verify that
(\ref{explicit})
implies a much more transparent relation 
${\rm vrad}\big({\calm}_d^{\rm tot} \big)= e^{-1/4}d^{-1/2}(1\pm O(d^{-1}))$ 
as $d\rightarrow \infty$, and similar two-sided estimates valid for all $d$. 

\medskip
This point of view allows to present in a compact way the gist of our results.
We start by listing, in Table  \ref{tab2},  bounds and asymptotics for volume radii
 of bases of various cones of maps acting on $N$--level density matrices.
Observe that the bounds for  volume radii of three middle sets (${\cal D}$, 
${\cal CP}$ and ${\cal T}$)  do not depend on dimensionality. 
On the other hand, the volume radii of the base for the
largest set  ${\cal P}$ of positive maps grow as $\sqrt{N}$,
while the volume radii of the smallest set ${\cal SP}$ of
superpositive maps decrease as $1/{\sqrt N}$.

\begin{table}[ht]
\caption{Volume radii for the bases 
      of mutually nested cones of positive maps
      which act on $N$--level density matrices. Here
      $r_{\rm CP}$ denotes the volume radius of 
      the base ${\cal CP}_N^{\rm b}$ of the set 
      of completely positive maps. The last column
    characterizes the  asymptotical properties,
    where $ r_{\cal X}^{\rm lim}  :=  \lim_{N\to \infty} {\rm vrad}({\cal
X}_N^b)$
   with ${\cal X}$ standing for ${\cal D}$, ${\cal CP}$ or ${\cal T}$.
   It is tacitly assumed that the limits exist, which we do not know for 
   ${\cal X} \neq {\cal CP}$ (the rigorous statements would involve then 
   $\liminf$ or $\limsup$, cf. Theorem \ref{thm2}). 
   The question marks ``?" indicate that we do not have asymptotic information 
   that is more precise than the one implied by the bounds in the middle
column. It is an interesting open problem whether $ r_{\cal T}^{\rm lim}$
admits a nontrivial (i.e., $<1$) {\em upper} bound; 
cf. remark (c) following Theorem \ref{thm2}.
      }
  
   \bigskip
{\renewcommand{\arraystretch}{1.67}
\begin{tabular}
[c]{|l| c  | c|}\hline \hline
{\quad \hskip 0.32cm Sets of maps \hskip 1.02cm} & 
{\hskip 1.00cm} \  Bounds for volume radii {\hskip 1.73cm}  & 

Asymptotics  {\hskip -0.04cm} \\
\hline
\end{tabular}
\\
\begin{tabular}
[c]{|c| c c c c c |c|} 
positive ${\cal P}$ & $\frac{1}{4}\sqrt{N} $ & $\le $
 & ${\rm vrad} ({\cal P}_N^{\rm b})$ & $\le $ & $6 \sqrt{N}$  &  ? \\
decomposable  ${\cal D}$ & $r_{\cal CP}$ & $\le $ 
 & ${\rm vrad} ({\cal D}_N^{\rm b})$ & $\le $ & $8 r_{\cal CP}$ &  $r_{\cal
D}^{\rm lim} \le 2$ \\
completely positive ${\cal CP}$ & $\frac{1}{2}$ & $ \le  $
 & $r_{\cal CP}:={\rm vrad} ({\cal CP}_N^{\rm b})$ & $ \le $ & $1$&  $r_{\cal
CP}^{\rm lim} \ =  \ e^{-1/4} $  \\

 {PPT inducing} ${\cal T}$ & $\frac{1}{4} r_{\cal CP}$ & $\le$  
 & ${\rm vrad} ({\cal T}_N^{\rm b})$ & $\le $ &  $r_{\cal CP}$ &  $r_{\cal
T}^{\rm lim} \ge \frac{1}{2} $ \\
super  positive ${\cal SP}$ & $\frac{1}{6} \frac{1}{\sqrt{N}} $ & $\le $
 & ${\rm vrad} ({\cal SP}_N^{\rm b})$ & $\le $ & $4 \frac{1}{\sqrt{N}}$ & ?
 \\
 \hline \hline
\end{tabular}
}
\label{tab2}
\end{table} 

 \bigskip 
 
The base of the set of completely positive maps 
acting on density matrices of size $N$ is up to a
rescaling by the factor $1/N$ equivalent to the set
of mixed states ${\cal M}_d^{\rm tot}$
of dimensionality $d=N^2$, and similarly for other 
cones of maps -- see eq. (\ref{isom1})-(\ref{isom3}).
Therefore, the results implicit in the last three rows of Table  \ref{tab2} are 
equivalent to the following bounds, presented in Table  \ref{tab3}, 
for the volume radii of the set of
quantum states and its subsets, 
some of which were known.

\begin{table}[ht]
\caption{Volume radii for the set of states 
  ${\cal M}_d^{\rm tot}$ of size $d$ and its subsets 
   ${\cal M}_d^{\rm PPT}$ and  ${\cal M}_d^{\rm sep}$.
   The latter two sets are well defined if the dimensionality
    $d$ is a square of an integer.
    Here $a \sim b$ means that $\lim_{d \raw \infty} a/b = 1$, while 
    $a \gtrsim b$ stands for $\liminf_{d \raw \infty} a/b \geq 1$. 
         }
  
    \bigskip
{\renewcommand{\arraystretch}{1.67}
\begin{tabular}
[c]{|l| c  | c|}\hline \hline
{\ \  Sets of states \hskip 0.05cm} & 
{\hskip 1.35cm} \  Bounds for volume radii {\hskip 2.3cm}  & 

\  Asymptotics  \hskip 1.3cm \\
\hline
\end{tabular}
\\
\begin{tabular}
[c]{|c| c c c c c |c|} 

all states & $\frac{1}{2} \frac{1}{\sqrt{d}}$ & $ \le  $
 & $r_{\rm tot}:={\rm vrad} ({\cal M}_d^{\rm tot})$ &
  $ \le $ & $\frac{1}{\sqrt{d}}$ & 
 $ r_{\rm tot} \sim e^{-1/4} \frac{1}{\sqrt{d}}$  \\
PPT states & $\frac{1}{4}\frac{1}{\sqrt{d}} r_{\rm tot}$ & $\le$  
 & $r_{_{\rm PPT}}  := {\rm vrad} ({\cal M}_d^{\rm PPT})$ & $\le $ & 
$\frac{1}{\sqrt{d}} r_{\rm tot}$ &  
 $r_{_{\rm PPT}}   \gtrsim \frac{1}{2} \frac{1}{\sqrt{d}}$ \\
separable states & $\frac{1}{6} \frac{1}{d} $ & $\le $
 & ${\rm vrad} ({\cal M}_d^{\rm sep})$ & $\le $ & $4 \frac{1}{d}$ & ?
 \\
 \hline \hline
\end{tabular}
}
\label{tab3}
\end{table} 

\medskip
Finally, we list  in Table \ref{tab4} the volume radii of the main objects of study in this paper:
the set ${\cal CP}_N^{\rm TP}$
of quantum operations and of other ``ensembles" of  trace preserving maps.
Each of these sets forms a $N^4-N^2$ cross-section of the corresponding 
$N^4-1$-dimensional base  (i.e., of ${\cal CP}_N^{\rm b}$ etc.). 

Although the volume of the larger set is sometimes known (\ref{explicit}), 
the cross-sections appear much harder to analyze.
Our approach does not aim at producing exact values
(even though here and in the previous tables we made an effort to obtain 
``reasonable" values for the numerical constants appearing in the
formulae).
Instead, we produce two-sided estimates  for the volume radius
of  ${\cal CP}_N^{\rm TP}$, which are quite tight in the asymptotic sense
(as the dimension increases) and analogous bounds for the sets of 
positive, decomposable, PPT--inducing and super--positive
trace preserving maps. 
Note that these  bounds are similar to
the results for the bases of all five sets presented  in Table \ref{tab2},
but are not their formal consequences.

\begin{table}[ht]
\caption{Asymptotic properties of volume radii for five
    nested sets of trace preserving maps.     Same caveat as in Table 2 applies
    to the limits in the second column.
 Upper and lower bounds  valid for all
$N$ (as in the middle columns of Tables  \ref{tab2} and  \ref{tab3}) can be likewise obtained.
}
  \bigskip \smallskip
  
{\renewcommand{\arraystretch}{1.67}
\begin{tabular}
[c]{|l| c  |}\hline \hline
\parbox{4.2cm}{\centering Sets of trace \\  preserving maps}

& {\hskip 0.6cm} \  asymptotics of their volume radii {\hskip 0.75cm}  \\
\hline
\end{tabular}
 \\
\begin{tabular}
[c]{|c| c c c c c |} 
positive ${\cal P}$ & $\frac{1}{4}$ & $\le $
 & $\lim_{N\to \infty}\frac{{\rm vrad} ({\cal P}_N^{\rm TP})}{\sqrt{N}}$ & $\le
$ & $6 $ \\
decomposable  ${\cal D}$ &  $e^{-1/4}$  & $\le $ 
 & $\lim_{N\to \infty}{\rm vrad} ({\cal D}_N^{\rm TP})$ & $\le $ & $ 2$ \\
completely positive ${\cal CP}$     &    & 
 & $\lim_{N\to \infty}{\rm vrad} ({\cal CP}_N^{\rm TP})$ & $= $ & $e^{-1/4}$ \\

 {PPT inducing} ${\cal T}$ &  $\frac{1}{2}$ & $\le$  
 & $\lim_{N\to \infty}{\rm vrad} ({\cal T}_N^{\rm TP})$ & $\le $ &  $e^{-1/4}$
\\
super  positive ${\cal SP}$ & $\frac{1}{6}  $ & $\le $
 & $\lim_{N\to \infty}\frac{{\rm vrad} ({\cal SP}_N^{\rm TP})}{1/\sqrt{N}}$ &
$\le $ & $4 $
 \\
 \hline \hline
\end{tabular}
}
\label{tab4}
\end{table} 

\medskip
While we concentrate in this work on the study of various classes of 
trace preserving maps, our approach allows deriving estimates of 
comparable degree of precision for other sets of quantum maps.
As an illustration, we sketch in Appendix \ref{appE} an argument 
giving tight bounds for the volume of {\em trace non-increasing} 
(TNI) maps. An exact formula for that volume was recently found by a different
method \cite{CSZ07} independently from the present work.  

Finally, let us point out that formula (\ref{explicit}) is valid only in the case when 
the underlying Hilbert space is complex, and that our analysis focuses 
on the complex setting,  as it is the one that is of immediate physical interest.
However, all the discussion preceding (\ref{explicit}) can be carried out 
also for real Hilbert spaces, and virtually all results that follow do have 
real analogues. This is because even when closed formulae are not 
available, the methods of geometric functional analysis allow to derive  
two-sided dimension free bounds on volume radii and similar parameters.
Accordingly, while in the real case one may be unable to precisely calculate 
coefficients such as $e^{-1/4}$ above,  it will be generally possible to 
%to deduce the asymptotic order of the relevant quantities as the dimension
%increases, and 
determine the relevant quantities  up to universal multiplicative constants.

%%%%%%%%%%%%
\newpage {\mbox{ }}
%\newpage
\subsection{A generic positive map acting on a high dimensional system
         is not decomposable}
         
         This is immediate from Table \ref{tab4}: the volume radius of the set of 
         positive trace preserving maps acting on an $N$ dimensional system is 
         of order $\sqrt{N}$,
         while the volume radius  of the corresponding set of decomposable
         trace preserving maps  is $O(1)$. Thus, for large $N$, the latter set 
         constitutes a very small part of the former one. Note that in order 
         to compare volumes we need to raise the ratio of the volume radii 
         to the power $N^4-N^2$, which yields roughly $N^{-N^4/2}$, a fraction 
         that is (strictly) subexponential in the dimension of the set.

\newpage {\mbox{ }}

%\newpage
\section{Known and preliminary results} \label{prelim}
\subsection{ Duality of cones }  

Spaces of operators or matrices are endowed with the canonical Hilbert-Schmidt 
inner product structure. The Choi-Jamio{\l}kowski isomorphisms transfers this 
structure to the space of quantum maps. We define
$$(\Phi,\Psi) := \bra D_\Phi, D_\Psi \ket_{HS}:=\Tr\, D_\Phi^{\dg} D_\Psi.$$
The spaces in question and the corresponding inner products are {\em a priori} 
complex.  However, if we restrict our attention to the {\em real} vector spaces 
of Hermicity-preserving maps $\Phi$ and Hermitian matrices $D_\Phi$,
which we will do in what follows, 
the scalar product becomes real and we may simply write
$$(\Phi,\Psi) =\Tr \, D_\Phi D_\Psi .$$

We next define a duality $^*$ for cones of maps via their representation
(or dynamical) matrix by
\begin{equation} \label{dualconeofmaps}
\mathcal{C}^*:=\{\Psi:\mathcal{M}_N \raw \mathcal{M}_N:
(\Phi,\Psi) \geq 0
\ \mbox{for\  all}
\ \Phi \in
\mathcal{C}\} .
\end{equation}
\noi 
This is a very special case of associating to a cone in a vector space 
the dual cone in the dual space (here $\mathcal{M}_d$ is identified with its
dual 
via the inner product $\bra \cdot, \cdot \ket_{HS}$).
Duality for cones of matrices and cones of maps is the same by
definition.

We point out that all the cones $\mathcal{C}$ we consider are non-degenerate,
i.e.,  they are of full dimension in the real vector space
$\mathcal{M}_{N^2}^{\rm sa}$ of Hermitian
matrices,  or in the space of linear maps commuting with $^\dg$ (equivalently,
every
map/matrix -- Hermicity-preserving or Hermitian, as appropriate --
can be written as the difference of two elements of $\mathcal{C}$)
and further $-\mathcal{C} \cap \mathcal{C} = \{0\}$.
Consequently, their duals are also non-degenerate.

\vskip 3mm

Since the cone of positive semi-definite matrices is self-dual, it follows that  
\begin{equation} \label{dualityCP}
\mathcal{CP}^*=\mathcal{CP} .
\end{equation}

\noi The superpositive cone $\mathcal{SP}$ may be defined via duality
\begin{equation} 
\mathcal{SP} := \mathcal{P}^* .  \label{dualityA}
\end{equation}
 By the bipolar theorem
for cones  ($(\mathcal{C}^*)^*=\mathcal{C}$), we then have 
\begin{equation}
\mathcal{SP}^*=\mathcal{P}. \label{dualityBZ}
\end{equation}
 [Note that the
bipolar theorem for {\em closed} cones follows, for example, from the easily
verifiable identity
$\mathcal{C}^* = -\mathcal{C}^\circ$, where $^\circ$ is the standard
polar defined by $K^\circ = \{x : \bra x, y \ket~\leq~1 \mbox{ for all } y \in
K \}$, 
and from the bipolar theorem for the standard polar, i.e., from the equality 
$(K^\circ)^\circ =K$ valid whenever $K$ is a closed convex set containing $0$.] 
Clearly  
$$\mathcal{SP} \subset \mathcal{CP} \subset \mathcal{P} ,$$
see Figure 1. 

\begin{figure}[htbp]
  \centering
  \includegraphics
  [width=.95\textwidth{}]{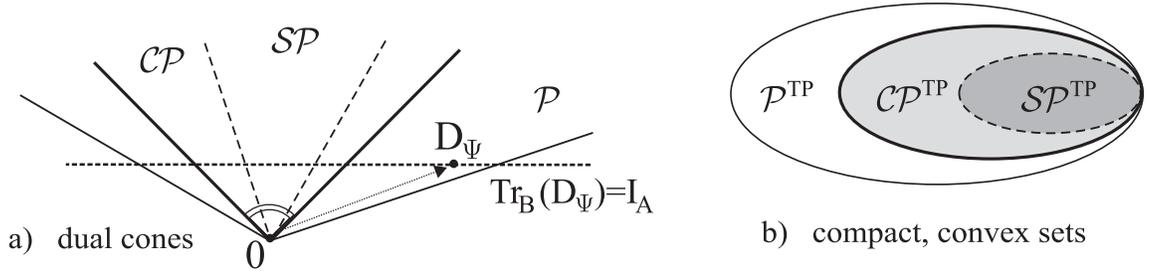}

  \caption{Sketch of sets of maps.
    a) The cone  $\cal P $ of positive maps 
   includes the cone $\cal CP $ of completely positive maps
   and its subcone $\cal CP $ containing the superpositive maps,
   dual to  $\cal P$.
   Trace preserving maps belong to the 
   cross--section of the cones with an affine plane of dimension 
   $N^4-N^2$ (and of codimension $N^2$), 
   representing the condition Tr$_B D={I}_A$.
   b) The sets of trace preserving maps in another perspective. 
   This is a complete picture for  $N=2$ since some of the cones 
   coincide, namely ${\mathcal P} =  {\mathcal D}$ 
   and ${\mathcal T} =  {\mathcal SP}$.   For $N \geq 3$,
   the complete picture is more complicated, see Figure 3. }
\end{figure}

\medskip
To clarify the duality relations  (\ref{dualityA}), (\ref{dualityBZ}) 
and the structure of the cone $\mathcal{SP}$,  we
recall that $\Phi$ is positive iff $D_\Phi$ is block positive,
which -- by definition --  is equivalent to $\Phi(\rho_\xi) \geq
0$ for every matrix of the form $\rho_\xi :=|\xi\ket \bra \xi|$,
that is, for every rank one positive semi-definite matrix.
In other words, for any $\xi \in \calh_A$ 
 and for any $\eta  \in \calh_B$,
$$
0 \leq \bra \Phi(|\xi\ket \bra \xi|) \eta, \eta \ket _{HS}
= \Tr \, \Phi(|\xi\ket \bra \xi|)\, |\eta\ket \bra \eta|
= \Tr \, D_\Phi (\rho_\xi \otimes\rho_\eta)
= \bra D_\Phi , \rho_\xi \otimes\rho_\eta\ket_{HS}
$$
where the first tracing takes place in $\calh_B$ (or $\calm_{N}$)
and the other 
in $\calh_A \otimes \calh_B$, or in $\calm_{N^2}$
(and similarly for the two Hilbert-Schmidt scalar products). 
This is the same as saying that $D_\Phi$ belongs to the 
cone of matrices that is dual to the separable cone 
(the cone generated by all $\rho_\xi \otimes \rho_\eta = \rho_{\xi \otimes \eta}$
or,
equivalently, by all products $\rho_A\otimes \rho_B$ of positive 
semi-definite matrices). By the bipolar theorem for cones, 
this is equivalent to the cone $\{D_\Phi \; : \; \Phi \in \mathcal{SP}\}$ 
being {\em exactly} the separable cone.

An alternative description of $\mathcal{SP}$, which justifies the 
``entanglement breaking" terminology,  is as follows: 
$\Phi$ is superpositive iff for every $k$ the extended quantum map 
$\Phi \otimes {\mathcal I}_k$ maps positive semi-definitive matrices
to (positive semi-definite) separable matrices, or states to separable states if $\Phi$ is 
trace preserving.

Sometimes (see, e.g., Appendix \ref{appB}) it is useful to work with extended
sets of maps such as the convex hulls of
${\mathcal P}_N^{\rm TP} \cup \{ 0\}$  or ${\mathcal P}_N^{{\rm b}} \cup \{ 0\}$.
For technical reasons, we find the latter one more useful; we will 
denote it by ${\mathcal P}^{\rm E}={\mathcal P}_N^{\rm E}$, and similarly for other cones.
Here $0$ denotes the ``zero" map,
which may be chosen as a reference point.
Further, one may consider symmetrized sets such as 
${\mathcal {CP}}^{\rm sym}={\mathcal {CP}}_N^{\rm sym}$, the
convex hull  of $-{\mathcal {CP}}^{\rm b} \cup {\mathcal {CP}}^{\rm b} $, where 
$-{\mathcal {CP}}^{\rm b}$ is the symmetric image of ${\mathcal {CP}}^{\rm b}$
with respect to $0$.  (Note that ${\mathcal {CP}}^{\rm sym}$ is also the convex 
hull of $-{\mathcal CP}^{\rm E} \cup {\mathcal CP}^{\rm E}$, see Figure 2 below.)
The advantage in using $0$-symmetric sets is that, first,
 they often admit an interpretation as unit balls with respect to natural norms 
 and, second, that symmetric convex bodies have been studied more 
 extensively than  general ones convex bodies.

\begin{figure}[htbp]
  \centering
  \includegraphics[width=0.40\textwidth{}]{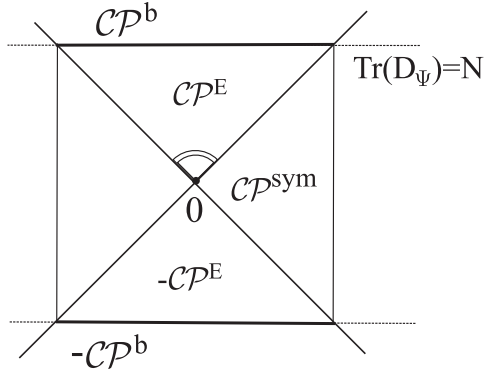}

  \caption{The set $\mathcal {CP}^{\rm b}$ of normalized quantum maps 
       arises as a cross-section of the unbounded cone of CP maps with 
  the hyperplane representing the condition Tr$\, D=N$.
  The set  ${\mathcal {CP}}^E$ of maps extended by the zero map is the convex
hull of 
${\mathcal CP}^{\rm b} \cup \{ 0\}$, 
  while ${\mathcal {CP}}^{\rm sym}$ is the symmetrized set,  
the convex hull  of $-{\mathcal {CP}}^{\rm b} \cup {\mathcal {CP}}^{\rm b} $. 
${\mathcal {CP}}^{\rm sym}$ may be identified with a ball in trace class norm,
whose radius equals $N$.
Analogous notation (and similar identifications)  may be employed 
for other sets of maps including
${\cal P, \, SP}$ etc., or for abstract cones.  }
\end{figure}

\vskip 2mm
We next introduce the auxiliary cone of completely co-positive  (CcP) maps
 
$$\mathcal{C}c \mathcal{P}=\{\Phi  : T \circ \Phi \in \mathcal{CP}\},$$
where  $T : \mathcal{M}_N \raw \mathcal{M}_N$ is the transposition map 
(which is positive, but not completely positive for $N>1$),
and the cone 
\begin{equation} \label{coneT}
\mathcal{T} :=  \mathcal{CP} \cap \mathcal{C}c \mathcal{P}  .
\end{equation}

\noi In terms of dynamical (Choi)  matrices,  
$D_{T\circ \Phi}$ is obtained from $D_{\Phi}$ by
transposing each block, i.e., by the partial transpose in the second system.
This means that  $\{D_{\Phi}\; : \; \Phi \in \mathcal{T}\}$  is exactly
$\mathcal{PPT}$,
the {\em positive  partial transpose cone} (positive semi-definite matrices whose 
partial transpose is also positive semi-definite). Since, as is easy to check, 
separable matrices  are in  $\mathcal{PPT}$, it follows that 
\begin{equation} \label{sptcp}
\mathcal{SP} 
\subset \mathcal{T} \subset \mathcal{CP} 
\end{equation}
For $N=2$ the sets  $\mathcal{T}$ and $\cal SP$
coincide, while for larger dimensions
the inclusion 
${\cal SP} \subset \mathcal{T}$ is proper 
as shown in Figure 3.  

\begin{figure}[htbp]
  \centering
\includegraphics[width=0.95\textwidth{}]{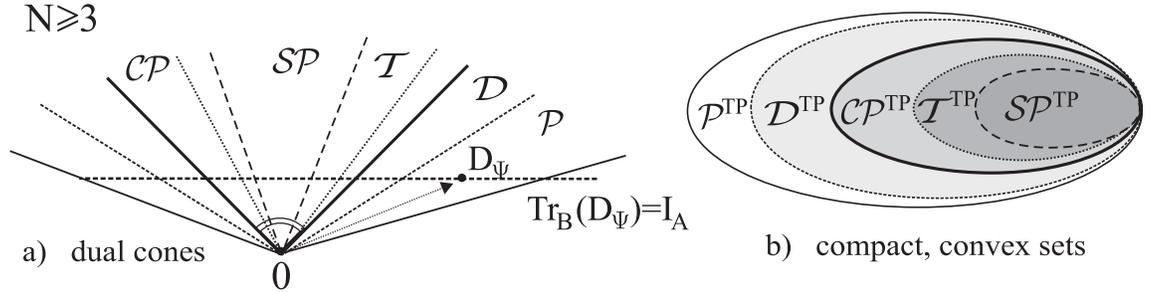}

\caption{Sketch of sets of maps for $N\geq3$.
    a) The cone  $\mathcal P $ of positive maps 
   includes a sequence of nested subcones:
  the cone  ${\mathcal D}$  of decomposable maps,
  ${\mathcal CP} $ of completely positive maps,
 $\mathcal{T}$ of maps which 
extended by identity impose positive partial transpose, 
and the cone $\mathcal SP $ of superpositive maps.
b) The sequence of nested subsets of the compact set
of positive trace preserving maps.}

\end{figure}
\vskip 3mm

Similarly to superpositive maps, there is an 
alternative description of $\mathcal{T}$
in the language of extended quantum maps: $\Phi \in \mathcal{T}$ iff
$\Phi \otimes {\mathcal I}_k$
is {\sl PPT inducing}  for any size $k$ of the extension, i.e.,
for any state $\rho$ acting on the bipartite system
its image, $\rho'=\Phi \otimes {\mathcal I}(\rho) \in \mathcal{PPT}$.
[The necessity of the latter condition follows by noticing that 
the partial transpose of $\rho'$ equals 
$(T \otimes {\mathcal I})\rho'=
(T\circ \Phi \otimes {\mathcal I})\rho$, which is positive semidefinite 
due to $T \circ \Phi$ being CP.]

A  quantum map $\Phi$ is called {\it decomposable}, if it may be expressed as
a sum of a CP map $\Psi_1$ and a 
another CP  map $\Psi_2$ composed with the transposition $T$, 
\begin{equation}
\Phi=\Psi_1 + T\circ  \Psi_{2} 
\label{decompos}
\end{equation}
or, equivalently, as a sum of a CP map and a CcP map. In other words, the 
cone $\mathcal{D}$ of decomposable maps is defined by 
$$
\mathcal{D}:= {\mathcal {CP}} + {\mathcal{C}c \mathcal{P}} 
$$
(the Minkowski sum).
Since the transposition preserves positivity, $\mathcal{D} \subset
\mathcal{P}$.
It is known \cite{St63,Wo76a}
that every one-qubit 
positive  map is decomposable,
so the sets ${\mathcal  P}_2$ and 
$\mathcal{D}_2$ coincide.
However, already for $N=3$
there exist positive, non--decomposable maps \cite{Cho75},
so $\mathcal{D}_3$ forms a proper subset of  ${\mathcal  P}_3$ -- see
Figure 3.

\medskip
It  follows from the identity $(\Phi,T \circ \Psi) = (T \circ \Phi,\Psi)$ 
valid for all $\Phi,  \Psi$ that
$$\mathcal{C}c \mathcal{P}^*=\mathcal{C}c \mathcal{P} .$$
 Accordingly, the dual cone $\mathcal{D}^*$ verifies 
\begin{equation} \label{pptstar}
\mathcal{D}^* 
= ({\mathcal {CP}} + {\mathcal{C}c \mathcal{P}} )^*
=\mathcal{CP}^* \cap {\mathcal{C}c \mathcal{P}}^*
={\mathcal {CP}} \cap {\mathcal{C}c \mathcal{P}}
=\mathcal{T}
\end{equation}
This is a special case of the identity 
$(\mathcal{C}_1 + \mathcal{C}_2)^* = 
\mathcal{C}_1^* \cap \mathcal{C}_2^*$ (the Minkowski sum) valid for any two 
convex cones
$\mathcal{C}_1, \mathcal{C}_2$.  It now follows by the bipolar theorem that 
\begin{equation} \label{pptstar2}
\mathcal{D}=\mathcal{T}^*.
\end{equation}

\noi
As
$\mathcal{SP}  \subset \mathcal{T} \subset \mathcal{CP}$ by (\ref{sptcp}), 
it follows by duality that
$$
 \mathcal{CP} \subset \mathcal{D} \subset  \mathcal{P} .
$$

\subsection{ Bases of cones and duality; the inradii and the outradii. \\
The symmetrized sets} \vskip 2mm 
\noi
We now return to the analysis of {\em bases} of cones of matrices, as defined
in section \ref{bases}.
As was to be expected, natural set-theoretic and algebraic operations on cones
induce 
analogous operations on bases of cones.
Sometimes this is trivial as in $(\calc_1 \cap \calc_2)^{\rm b} = \calc_1^{\rm
b} \cap \calc_2^{\rm b}$,
in other cases simple:  $(\calc_1 + \calc_2)^{\rm b} = \co (\calc_1^{\rm b}
\cup \calc_2^{\rm b})$, 
where $\co$ stands for the convex hull.
% (the latter follows by renormalization of the summands in $\calc_1 + 
%   \calc_2$).
What is more interesting and somewhat surprising is that also duality of cones 
carries over to precise duality of bases in the following sense.

\begin{lemma} \label{baseduality}
Let $\calv$ be a real Hilbert space, $\calc \subset \calv$ a closed convex cone
and let $e \in \calv$ 
be a unit vector such that $e \in \calc \cap \calc^*$.  Set
$V^{\rm b} := \{x \in \calv \; : \; \bra x,e \ket =1\}$ and let and $\calc^{\rm
b} = \calc \cap V^{\rm b}$ and 
 $(\calc^*)^{\rm b} = \calc^* \cap V^{\rm b}$ be the corresponding bases of
$\calc$ and $\calc^*$. 
Then
\begin{equation} \label{base}
(\calc^*)^{\rm b} := \calc^* \cap V^{\rm b} = \{y \in V^{\rm b} \; : \; \forall
x \in \calc^{\rm b} \ \ \bra -(y-e),x-e \ket \leq1\} .
\end{equation}
In other words, if we think of  $V^{\rm b}$ as a vector space with the origin
at $e$, and of $\calc^{\rm b}$ and  $(\calc^*)^{\rm b}$  as subsets of that
vector space, then  $(\calc^*)^{\rm b} = -(\calc^{\rm b})^\circ$. 
\end{lemma}
Recall that for abstract cones  $\calc \subset \calv$, the dual cone  $\calc^*$ 
is defined (cf. (\ref{dualconeofmaps})) via
$$
\calc^* := \{x \in \calv \; :\; \forall \, y \in \calc \ \bra x, y \ket \geq 0\}.
$$
This elementary Lemma %\ref{baseduality} 
seems to be a folklore result, but does not appear in
standard references 
for convexity (the best source we were pointed to after consulting specialists
was Exercise 6, \S 3.4  of \cite{grunbaum}). However, once stated, the Lemma
is straightforward to prove. 
If $\bra x,e \ket= \bra y,e \ket= 1$, then $\bra -(y-e),x-e \ket = -\bra
y,x\ket +1$ and so the condition from (\ref{base}) can be restated as 
$$
\forall x \in \calc^{\rm b} \ -\bra y,x\ket +1 \leq 1 \ \ \Leftrightarrow \ \
\forall x \in \calc^{\rm b} \ \bra y,x\ket \geq 0 .
$$
Since under our hypotheses $\calc^{\rm b}$ generates $\calc$, the latter
condition is equivalent to 
$\bra y,x\ket \geq 0$ for all $x \in \calc$, i.e., to $y \in \calc^*$, as
required.

\vskip 3mm
Let us now return to our more concrete setting of $\calv=\mathcal{M}_d^{\rm
sa}$ 
(endowed with the Hilbert-Schmidt scalar product) and 
$e = {I}_d/d^{1/2}$. 
Even more specifically, we will consider
 $\calv=\mathcal{M}_{N^2}^{\rm sa}$, identified via the Choi-Jamio{\l}kowski
isomorphism 
with the space of Hermicity preserving quantum maps on $\mathcal{M}_N$, 
and the cones that we defined in prior section. Note that  the quantum map 
associated to  $e = { I}_{N^2}/N$ is the so-called 
``completely depolarising map," which is usually denoted by $\Phi_*$ and whose
action is  
described by $\Phi_*(M) = (N^{-1} \Tr \, M ) \,{ I}_{N}$.  The 
duality relations for cones (\ref{dualityCP}), (\ref{dualityA}),
(\ref{dualityBZ}) and 
(\ref{pptstar}), (\ref{pptstar2}) combined with Lemma  \ref{baseduality} imply
now 

\begin{cor} We have the following duality 
relations for the bases of cones
$$
(\mathcal{CP}^{\rm b})^\circ = -\mathcal{CP}^{\rm b}, \ 
(\mathcal{SP}^{\rm b})^\circ  = -\mathcal{P}^{\rm b},
\ 
(\mathcal{P}^{\rm b})^\circ  = -\mathcal{SP}^{\rm b}$$ 
\begin{equation} 
\label{dualbase}
(\mathcal{D}^{\rm b})^\circ  = -\mathcal{T}^{\rm b},
\ 
(\mathcal{T}^{\rm b})^\circ  = -\mathcal{D}^{\rm b} ,
\end{equation}
where both the polarity and the negative signs refer to the vector structure in 
$H^{\rm b} = \{ \Phi \; : \; \Tr\, D_\Phi = \Tr \,\Phi( {I }_N)=N\}$
with $\Phi_*$ as the origin.
\end{cor}
In other words, we have for example 
$$
\mathcal{D}^{\rm b} = \{\Phi \in H^{\rm b} \ : \  \forall \Psi \in
\mathcal{T}^{\rm b} \ 
(-(\Phi -\Phi_*), (\Psi - \Phi_*)) \leq 1\} .
$$
%
%\smallskip
While the duality relations {\em for cones} described in the preceding
subsection 
are rather well known, the duality  {\em for bases} in the present generality 
appears to be a new observation. When combined with standard results 
from convex geometry, most notably Santal\'o and inverse Santal\'o inequalities
\cite{Sa,BM} (see below),
and other tools of geometric functional analysis, it allows for relating 
volumes of bases of cones to those of the dual cones, and ultimately 
for asymptotically precise estimates  of these volumes and of volumes of 
the  corresponding sets of trace preserving maps.

\smallskip
Let us also note here one immediate but interesting (and presumably known) 
consequence of the duality  relations.
\begin{cor} \label{radii}
For each of the sets $\mathcal{CP}^{\rm b}_N, \mathcal{SP}^{\rm b}_N,
\mathcal{P}^{\rm b}_N, \mathcal{D}^{\rm b}_N$
and $\mathcal{T}^{\rm b}_N$, the Euclidean (i.e., Hilbert-Schmidt) in-radius is
$(N^2-1)^{-1/2}$ and the Euclidean out-radius is $(N^2-1)^{1/2}$.
\end{cor}
We observe first that, for each of the above sets, $\Phi_*$ is the only element
that is invariant under isometries of the set. Accordingly, it is enough to
restrict attention to 
Hilbert-Schmidt balls centered at $\Phi_*$.  
For $\mathcal{CP}^{\rm b}_N$, the assertion is just a reflection of the
elementary fact that 
${\mathcal M}^{\rm tot}_{d}$ contains a Hilbert-Schmidt ball of radius
$1/\sqrt{d(d-1)}$
centered at the maximally mixed state $\rho_*$, and that the distance from 
$\rho_*$ to pure states is $\sqrt{1-1/d}$. For $\mathcal{SP}^{\rm b}_N$, it is
a
consequence of equality of in-radii of ${\mathcal M}^{\rm tot}_{N^2}$ and 
${\mathcal M}^{\rm sep}_{N^2}$ (in the bivariate case) established in \cite{GB}
(the out-radius of the latter is of course attained on pure separable states).
It then follows that the in- and out-radii must be the same for the
intermediate set
$\mathcal{T}^{\rm b}_N$. Finally, since the out-radius of $K^\circ$ is the
reciprocal
of the in-radius of $K$ (and {\em vice versa}), we deduce the assertion for 
$\mathcal{P}^{\rm b}_N$ and $\mathcal{D}^{\rm b}_N$ via (\ref{dualbase}). 

It is curious to note that the statement about the out-radius of
$\mathcal{P}^{\rm b}_N$
is equivalent -- via simple geometric arguments -- 
to the following fact (which {\em a posteriori} is true)

\medskip
{\em If $M = (M_{jk})_{j,k=1}^N $ is a block-positive matrix, then $\Tr\, (M^2)
\leq (\Tr \,M)^2$}.

\medskip \noindent
It would be nice to have a simple direct proof of the above inequality, as 
it would yield (via Lemma \ref{baseduality} and (\ref{dualbase})) an alternate
derivation of the result from  \cite{GB} concerning the in-radius of 
the set of separable states in the bivariate case.

\medskip
Similarly, the best (i.e., the smallest) constant $R$ in the inclusion
$$
\mathcal{CP}^{\rm b}-\Phi_* \subset R \, (\mathcal{SP}^{\rm b}-\Phi_*)
$$
is the same as the best constant in
$$
\mathcal{P}^{\rm b}-\Phi_* \subset R \, (\mathcal{CP}^{\rm b}-\Phi_*) .
$$
It has been shown in \cite{GB} that the optimal $R$ satisfies
$N^2/2+1 \leq R \leq N^2-1$. [The upper bound follows just from the
formulae for the inradius of  $\mathcal{SP}^{\rm b}$  and the outradius 
of $\mathcal{CP}^{\rm b}$ (or, equivalently, ${\mathcal M}^{\rm sep}$, 
$\mathcal{M}^{\rm tot}$).] 
Again, there could be a more direct elementary argument.

\begin{rem} \label{TPradii}
The Euclidean inradii and outradii of $\mathcal{CP}^{\rm TP}_N,
\mathcal{SP}^{\rm TP}_N, \mathcal{P}^{\rm TP}_N, \mathcal{D}^{\rm TP}_N$ and
$\mathcal{T}^{\rm TP}_N$ are the same as for the larger 
$\calc^{\rm b}$-type sets, i.e.,  $(N^2-1)^{-1/2}$ and  $(N^2-1)^{1/2}$.
\end{rem}
As pointed out in the arguments following the statement of Corollary
\ref{radii},
while the fact that the inradii and outradii of all sets in that Corollary 
are identical is nontrivial, there is no mystery about at least some of the
maps 
(or directions)  that witness them. In the language of the sets of states
(i.e., matrices with trace one normalization) such witnesses are, for outradii
 pure states, and universal witnesses that work 
 for all five sets are pure separable states.
 By duality (i.e., Lemma \ref{baseduality}), direction that witness inradii
 (for all sets) are obtained by reflecting a pure separable state with respect
to  
 the maximally mixed state $\rho_*$. In the language of quantum maps
 purity (i.e., the Choi matrix being of rank one) corresponds to the map 
 being of the form  $\rho \raw v^\dagger \rho v$ (Kraus rank one),
 and the trace preserving condition is then equivalent to $v$ being unitary.
 If that unitary is separable (i.e., a tensor product of two unitaries acting 
 on the first and second system), the corresponding pure state will 
 be separable. This means that universal witnesses of outradii of  
 $\calc^{\rm b}$-type sets exist also in the smaller set
 by the trace preserving condition (\ref{tpcond}), i.e., inside  the 
 $\calc^{\rm TP}$-type sets. Since condition (\ref{tpcond}) defines 
 an affine subspace, the ``opposite" directions giving witnesses 
 to the inradii also belong there.

\bigskip 
An alternative use of duality considerations involves symmetrized sets 
(cf. Figure 2).  If $\calc \subset \calv$ is a cone and $\calc^{\rm b}$ its
base, we define $\calc^{\rm sym} 
:= \co (-{\mathcal {C}}^{\rm b} \cup {\mathcal {C}}^{\rm b} )$; the minus sign
referring 
now to  the symmetric image with respect to $0$.  If, as earlier, $e$ is the 
distinguished point of $\calc \cap \calc^*$ defining  
${\mathcal {C}}^{\rm b}$ and $({\mathcal {C}}^*)^{\rm b}$,
then
\begin{equation} \label{order}
\left(\calc^{\rm sym}  \right)^\circ 
= ( e -{\mathcal{C}}^* ) \cap (-e+{\mathcal{C}}^*) ,
\end{equation}
where the polarity has now the standard meaning (i.e., inside the entire 
space $\calv$ {\em and} with respect to the origin).  In other words, 
the polar of $\calc^{\rm sym}$ is the order interval $[-e, e]$, in the sense of
the 
order induced by the cone ${\mathcal {C}}^*$.   The advantages of this approach 
is that we find ourselves in the category of centrally symmetric convex sets,
which is better understood than that of general convex sets, and that
frequently 
the object in question ($\calc^{\rm sym}$ and its polar)  have natural
functional analysis 
interpretation as balls in natural normed spaces. One disadvantage is that  
in place of one very simple operation (symmetric image with respect to $e$) 
we have two elementary and manageable, but somewhat non-trivial operations 
(symmetrization and passing to order intervals). 
We postpone the discussion of (\ref{order}) and related issues to the Appendix.

\subsection{Volume radii and duality:  Santal\'o and inverse Santal\'o
inequalities}

The classical Santal\'o inequality  \cite{Sa} asserts that 
if $K \subset \mathbb{R}^m$ is a $0$-symmetric convex body and $K^\circ$ its
polar body, then
${\vol (K)} \;{\vol (K^\circ)} \ \le \ \big({\vol \big(B_2^m\big)}\big)^2$ or,
in other words
\begin{equation}\label{santalo}
{\rm vrad (K)}\; {\rm vrad (K^\circ)} \leq 1 .
\end{equation}
Moreover, the inequality holds also for 
not-necessarily-symmetric convex sets after an appropriate translation, 
in particular if the origin is the centroid of $K$ or of $K^\circ$, a condition
that will be satisfied for all sets we will consider in what follow.
Even more interestingly, there is a converse inequality  \cite{BM}, usually 
called ``the inverse Santal\'o inequality,"
\begin{equation}\label{revsantalo}
{\rm vrad (K)}\; {\rm vrad (K^\circ)} \geq c  
\end{equation}
for some universal numerical constant $c>0$, independent of the convex body $K$ 
(symmetric or not) and, most notably, of 
its dimension $m$. 

The inequalities (\ref{santalo}), (\ref{revsantalo}) together imply that, 
under some natural hypotheses (which are verified in most 
of cases of interest), the volume radii of a convex body and of its polar are 
approximately (i.e., up to a multiplicative universal numerical constant) 
reciprocal. By Lemma \ref{baseduality}, the same is true for the base of
a cone and that of the dual cone.  This observation reduces, roughly by a
factor of 2,
the amount of work needed to determine the asymptotic behavior of volume 
radii of, say, sets from the third column of Table 1.  We note, however, that
since, 
at present, there are no good estimates for the constant $c$ from
(\ref{revsantalo}) 
if $K$ is not symmetric, it is often more efficient to revisit arguments from 
\cite{Sza04,AS} which allow to estimate volume radii of polar bodies 
without resorting to the inverse Santal\'o inequality. (An argument yielding 
reasonable value of $c$ for {\em symmetric} bodies was 
recently given in \cite{kuperberg}.)

\newpage
\section{Volume estimates:  precise statements and approximate arguments}
\label{results}

The results stated in section \ref{prelim} allow us, in combination with known
facts, to determine the asymptotic orders 
(as $N \raw \infty$) for  the volume radii  (and hence reasonable estimates for
the volumes) 
of bases of all cones of quantum maps discussed up to this point. 
%$\Pi^\circ$.
Our goal is slightly more ambitious; we want to find not 
just the asymptotic order of each quantity, but also establish inequalities
valid in every 
fixed dimension and involving explicit fairly sharp numerical constants. 
Specifically, we will show the following

\begin{thm} \label{thm2}
We have the following inequalities, valid for all $N$, and the following 
asymptotic relations

\medskip
\noi {\rm (i)} \ \ \;$
\frac 12 \leq \vrad\big(\mathcal{CP}_N^{\rm b}\big) \leq 1, 
\ \ \ \lim_{N \raw \infty} \vrad\big(\mathcal{CP}_N^{\rm b}\big) = e^{-1/4} \approx 0.779
$\\
{\rm (ii)}  \ \ $
\frac 14 N^{1/2}\leq \vrad\big(\mathcal{P}_N^{\rm b}\big) \leq 6N^{1/2} 
$\\
{\rm (iii)}  \ \,$
\frac 16 N^{-1/2} \leq \vrad\big(\mathcal{SP}_N^{\rm b}\big) \leq 4N^{-1/2} 
$\\
{\rm (iv)}  \ \ $
\frac 14 \leq \frac {\vrad\big(\mathcal{T}_N^{\rm
b}\big)}{\vrad\big(\mathcal{CP}_N^{\rm b}\big)} \leq 1, 
\ \ \ \frac{e^{1/4}}2 \leq \liminf_N   \frac {\vrad\big(\mathcal{T}_N^{\rm
b}\big)}{\vrad\big(\mathcal{CP}_N^{\rm b}\big)}  $\\
{\rm (v)}  \ \ \;$
1 \leq  \frac {\vrad\big(\mathcal{D}_N^{\rm
b}\big)}{\vrad\big(\mathcal{CP}_N^{\rm b}\big)}  \leq 8, 
\ \ \  \limsup_N   \frac {\vrad\big(\mathcal{D}_N^{\rm
b}\big)}{\vrad\big(\mathcal{CP}_N^{\rm b}\big)}  
\leq 2 e^{1/4}$
\end{thm}

\noi {\em Remarks} : (a) Estimates on volume radii listed in Table \ref{tab2}  are
either identical to the corresponding inequalities stated above, 
or follow by the same argument. \\(b) Since the asymptotic orders of the volume radii of the
families $\mathcal{CP}_N^{\rm b}$, $\mathcal{T}_N^{\rm b}$ and
$\mathcal{D}_N^{\rm b}$ are the same, we chose  -- for greater transparence --
to compare the volume radii of the two latter sets to that of
$\mathcal{CP}_N^{\rm b}$ in {\rm (iv)} and {\rm (v)}, rather than give
separate estimates for each of these quantities. \\
(c) It is an interesting open problem whether there exists a universal constant
$\alpha <1$ such that 
${\vrad\big(\mathcal{T}_N^{\rm b}\big)} \leq \alpha
\,{\vrad\big(\mathcal{CP}_N^{\rm b}\big)}$ for all $N>2$ or, equivalently, 
``is  
$\vol\big(\mathcal{M}^{\rm PPT}_{N^2}\big) \leq \alpha^{N^4} \,
\vol\big(\mathcal{M}^{\rm tot}_{N^2}\big)$
for some $\alpha <1$ and {\em all} $N>2$?" Analogous question may be asked about
comparing  
${\vrad\big(\mathcal{D}_N^{\rm b}\big)}$ and ${\vrad\big(\mathcal{CP}_N^{\rm
b}\big)}$. Inquiries to similar effect can be found in the literature 
\cite{ZHSL98, HHH01}.\\
(d) It is likely that the {\em asymptotic} bound  $2 e^{1/4}$ in  {\rm (v)}
holds actually for all $N$. Indeed, there is a strong numerical evidence that 
the estimate $\vrad\big(\mathcal{CP}_N^{\rm b}\big) \geq e^{-1/4}$ from {\rm
(i)} is valid for all $N$ and not just in the limit. (In view of the explicit
character of the formula 
(\ref{explicit}) this issue shouldn't be too difficult to resolve.) Should that
be the case, 
the next step would be to carefully analyze the dependence of
${\vrad\big(\mathcal{D}_N^{\rm b}\big)}$
on $N$ given by the arguments presented in this paper.

\medskip Since the bases of cones, whose volume radii are described by
Theorem \ref {thm2}, are effectively homothetic images, with ratio $N$, of the
corresponding sets of trace one matrices (see Table 1 and the formulae that
follow it),  some of the inequalities/relations of Theorem \ref{thm2} follow
from known estimates for the volumes of various sets of states, particularly
if we do not insist on 
obtaining ``good" numerical constants that are included in the statements.  
For example, the estimates in statement {\rm (iii)} are contained in Theorem 1
from \cite{AS}; one obtains the constants $\frac16$ and 4 by going over the
proof of that Theorem specified to bilateral systems.
Similarly, the statement  {\rm (iv)} is 
(essentially) a version of Theorem 4 from \cite{AS} which asserts that, in the
present language,  
$\vrad\big(\mathcal{M}^{\rm PPT}_{N^2}\big)/{\vrad\big(\mathcal{M}^{\rm
tot}_{N^2}\big)} \geq c_0$ 
for some constant $c_0>0$ independent of the dimension $N$ (the upper estimate
with constant 1 is trivial).
However, the argument from \cite{AS}   yields only $c_0=\frac18$ 
and  $\frac{e^{-1/4}}4$  for the asymptotic lower bound.

 Next, the asymptotic relation in {\rm (i)} follows from the explicit formula
(\ref{explicit}); see the comments following (\ref{explicit}). Presumably, the
estimates  in {\rm (i)} can also be derived from (\ref{explicit}), but there
are more elementary arguments. 
For a simple derivation of the lower bound from the classical  Rogers-Shephard
inequality \cite{RS} see \cite{Sza04}, section II.  And here is an apparently
new proof of the upper bound: combine the duality results of the preceding
section, specifically the identification $(\mathcal{CP}^{\rm b})^\circ  =
-\mathcal{CP}^{\rm b}$ from (\ref{dualbase}), with the Santal\'o inequality
(\ref{santalo}) to obtain 
$$
1 \geq \vrad\big(\mathcal{CP}^{\rm b}\big) \vrad\big((\mathcal{CP}^{\rm
b})^\circ \big) =
\vrad\big(\mathcal{CP}^{\rm b}\big)\vrad\big(-\mathcal{CP}^{\rm
b}\big)=\vrad\big(\mathcal{CP}^{\rm b}\big)^2,
$$
as required. We recall that, in the context of  (\ref{dualbase}), the
operations $^\circ$ and $-$ take place in the space $H^{\rm b}$ of quantum
maps verifying $\Tr\, D_\Phi = N$, with  $\Phi_*$  thought of as the origin;
note that $\Phi_*$ is the centroid of  $\mathcal{CP}^{\rm b}$ and so
(\ref{santalo})  with $K=\mathcal{CP}^{\rm b}$ indeed does apply in that
setting. 

\smallskip 
Arguments parallel to the last one lead to versions of the remaining statements
with {\em some} universal constants. For example, the identification
$(\mathcal{SP}^{\rm b})^\circ  = -\mathcal{P}^{\rm b}$ combined with the
Santal\'o inequality (\ref{santalo}) and its inverse (\ref{revsantalo}) leads
to
$$
1 \geq  \vrad\big(\mathcal{SP}^{\rm b}\big)\vrad\big(\mathcal{P}^{\rm b}\big)
\geq c,
$$
where $c$ is the (universal) constant from (\ref{revsantalo}). 
Combining the above inequality with {\rm (iii)} we obtain 
$\frac c4 \, N^{1/2} \leq \vrad\big(\mathcal{P}^{\rm b}_N\big) \leq 6N^{1/2}$.
Similarly, 
$1 \geq  \vrad\big(\mathcal{T}^{\rm b}\big)\vrad\big(\mathcal{D}^{\rm b}\big)
\geq c$ combined with 
(the already shown version of) {\rm (iv)} and with {\rm (i)} implies 
$\frac {\vrad\big(\mathcal{D}_N^{\rm b}\big)}{\vrad\big(\mathcal{CP}_N^{\rm
b}\big)}  \leq 32$ and 
$\limsup_N   \frac {\vrad\big(\mathcal{D}_N^{\rm
b}\big)}{\vrad\big(\mathcal{CP}_N^{\rm b}\big)}  
\leq 4{e^{3/4}}$.  
As the constants in Theorem  \ref {thm2} are not meant to be optimal, we
relegate the somewhat more involved (but still based on classical facts)
arguments yielding them to Appendix \ref{appA}.

\medskip
The inequalities of  Theorem  \ref {thm2} compare volumes of bases of cones,
that is, 
sets of maps $\Phi$ normalized by the condition that the trace of $D_\Phi$, the 
corresponding Choi (or dynamical) matrix, is N (or $\Tr \,\Phi({I}_N)=N $).
[Of course, any other normalization -- most notably $\Tr \,D_\Phi = 1$ leading
to sets of states --
would 
work just as well for comparing volumes provided we were consistent.] 
However, if we want to study quantum operations, i.e.,  trace-preserving
quantum maps
(or, similarly, unital maps), then -- as explained in the previous sections --
the
corresponding constraints are stronger than just normalization by trace: 
in each case we are looking at an $N^2$-codimensional section of the cone
as opposed to the 1-codimensional base.
However, in either case the codimension is much smaller than the dimension,
which is $N^4-N^2$.  The following technical result will imply that then, under 
relatively mild additional assumptions assuring that the base of the cone is 
reasonably balanced (which will be the case for all the cones we studied), 
the volume radius of the section will be very close to that of the entire base.

\begin{prop}\label{section}
Let $K$ be a convex body in an  $m$-dimensional Euclidean 
space with centroid at $a$, 
and let $H$ be a $k$-dimensional affine subspace 
passing through $a$.
Let  $r=r_K$ and $R=R_K$ be the in-radius and out-radius of $K$.
Then 
\begin{equation} \label{vradsection}
\left(\vrad(K) \ R^{-\frac{m-k}{m}} \ b(m,k)\right)^{\frac{m}{k}} \leq 
\vrad({K \cap H})
\leq \left(\vrad(K) \ r^{-\frac{m-k}{m}} \ b(m,k) {m \choose k}^{\frac
1m}\right)^{\frac{m}{k}},
\end{equation}
where
$b(m,k):=\left(\frac{\mbox{vol}_m(B^m_{2})}
{\mbox{vol}_k(B^k_{2})\mbox{vol}_{m-k}(B_2^{m-k})}
\right)^{\frac{1}{m}}$ .
\end{prop}

\medskip \noi 
The proof of the Proposition is relegated to Appendix \ref{appC}; now we 
explain its consequences. First, let us analyze the 
parameters that appear in (\ref{vradsection}).
By Corallary \ref{radii}, for all bases of cones that we consider here we have
$r = 1/\sqrt{d-1}$ and $R = \sqrt{d-1}$, where $d=N^2$.
Next, we have $m=d^2-1=N^4-1$, 
$k=d^2-d=N^4-N^2$ and $m-k=d-1=N^2-1$, in particular  
$\frac mk = 1 + \frac1{N^2}= 1+\frac 1d$
and $\frac{m-k}{m} = \frac 1{N^2+1} = \frac 1{d+1}$.
 Further, the quantity 
$b(m,k)=\left(\frac
{\Gamma(k/2+1)\Gamma((m-k)/2+1)}{\Gamma(m/2+1)}\right)^{\frac{1}{m}}$ 
(related to the Beta function) is easily shown to satisfy  
$1/\sqrt{2} < b(m,k) < 1$ (for our values of $m, k$ it is actually 
$1- O(\frac{\log{N}}{N^2})$). 
Similarly, $1 \leq  {m \choose k}^{\frac 1m}\ \leq 2$ for all $k, m$ 
and $1+ O(\frac{\log{N}}{N^2})$ for  our values of $m, k$.
Consequently, if $\vrad(K)$  is subexponential in $d$ 
(in our applications it is a low power of $N$, hence of $d$), 
then
$\vrad({K \cap H})/\vrad(K) \raw 1$ as $N \raw \infty$. 

This leads to 

\begin{thm} \label{thm5}
We have the following asymptotic relations

\medskip
\noi {\rm (i)} \ \ \;$
\lim_{N \raw \infty} \vrad\big(\mathcal{CP}_N^{\rm TP}\big) = e^{-1/4}
$\\
{\rm (ii)}  \ \ $
\frac 14 \leq \liminf_N \frac{\vrad\big(\mathcal{P}_N^{\rm TP}\big)}{N^{1/2}} 
\leq \limsup_N \frac{\vrad\big(\mathcal{P}_N^{\rm TP}\big)}{N^{1/2}}\leq 6 
$\\
{\rm (iii)}  \ \,$
\frac 16  \leq  \liminf_N \frac{\vrad\big(\mathcal{SP}_N^{\rm
TP}\big)}{N^{-1/2} }
\leq  \limsup_N \frac{\vrad\big(\mathcal{SP}_N^{\rm TP}\big)}{N^{-1/2} }\leq
4 $\\ 
{\rm (iv)}  \ \ $
\frac{e^{1/4}}2 \leq \liminf_N   \frac {\vrad\big(\mathcal{T}_N^{\rm
TP}\big)}{\vrad\big(\mathcal{CP}_N^{\rm b}\big)}  $\\
{\rm (v)}  \ \ \;$
 \limsup_N   \frac {\vrad\big(\mathcal{D}_N^{\rm
TP}\big)}{\vrad\big(\mathcal{CP}_N^{\rm b}\big)}  
\leq 2 e^{1/4}$
\end{thm}
Upper and lower bounds in the spirit of Theorem \ref{thm2} (i.e., valid for all
$N$) can be likewise obtained.

The reader may wonder why we perform our initial analysis 
on bases of cones rather  than working directly with the smaller 
sets of trace preserving maps. The reason for this is two-fold.
First, the bases being homothetic to various sets of states,
any information about them is at the same time more 
readily available and interesting by itself. 
Second, while we do have -- as a consequence of Lemma \ref{baseduality} --
nice duality relations between bases of cones, similar results 
for sets of trace preserving maps are just not true.
As a demonstration of that phenomenon we show in 
Appendix \ref{appD} that, in contrast to the bases $\mathcal{CP}^{\rm b}$,
the sets $\mathcal{CP}^{\rm TP}$ are very far from being self-dual
in the sense of (\ref{dualbase}).

 \section{Conclusions}
 
 We derived tight explicit bounds for the effective radius
(in the sense of Hilbert-Schmidt volume), or volume radius,  of the set of
quantum operations acting on density matrices of size $N$,
and for other convex sets of trace preserving maps acting
such matrices such as positive, decomposable, PPT inducing or 
 superpositive maps. The novelty of our approach depends on
systematic use of duality to derive quantitative estimates,
and on technical tools, some of which are not very familiar
even in convex analysis.
 
Since the volume radii of the sets of trace preserving maps that are positive 
 display a different dependendce on the dimensionality than those of 
the smaller set of  decomposable maps, 
the ratio of the volumes of the latter and the former set tends rapidly to $0$ 
as the dimension increases. In other words, a generic positive trace preserving 
map is not decomposable and, {\em a fortiori}, not completely positive.
Thus we were able to prove a stronger statement than
 the one advertised in the title of the paper.
Similarly, a generic PPT inducing quantum operation (and, {\em a fortiori}, 
a generic  quantum operation) is not  superpositive. 
Analogous relations (some of which were known) 
exist between  the sets of states related to those of maps via 
the Jamio{\l}kowski isomorphism.

\vskip.5cm \noindent 
{\em Acknowledgements}:
We enjoyed inspiring discussions with I.~Bengtsson, F.~Benatti, V.~Cappellini
and H--J.~Sommers. 
We acknowledge financial support from the Polish Ministry of
Science and Information Technology
under the grant DFG-SFB/38/2007, 
from the National Science Foundation (U.S.A.), 
and from the European Research Projects SCALA and PHD.

%%%%%%%%%%%%%%%%%%%%%%%%%%

\newpage \section{Appendices}
\subsection{Better constants in Theorem \ref{thm2}:  mean width, 
Urysohn inequality and related tools} \label{appA}

The arguments given in the preceding section did not yield 
the asserted values of the constant $\frac 14 $  in part (ii),  the constants
$\frac {e^{1/4}}2$ 
and $\frac 14 $  in part (iv), and the constants 8 and $2 e^{1/4}$  in part (v) 
of Theorem \ref{thm2}. We will now 
present the somewhat more involved 
line of reasoning that does yield these constants.

\smallskip 
The following concepts will be helpful in our analysis.
If $K \subset \mathbb{R}^m$ is a convex body containing 
the origin in its interior, one defines  the {\em gauge of}  $K$  via 
$$
\|x\|_K := \inf \{ t \geq 0 \, : \, x \in tK\}.
$$
Roughly, $\|x\|_K$ is the norm, for which $K$ is the unit ball,
except that there is no symmetry requirement.
Next, the {\em mean width} of $K$ 
(or, more precisely, the mean {\em half-width}) is defined by 
$$
w(K):=\int_{S^{m-1}} \|x\|_{K^\circ} dx
=\int_{S^{m-1}} \max_{y\in K} \bra x,y\ket dx
$$
(integration with respect to the normalized Lebesgue measure on $S^{m-1}$).
A classical result known as Urysohn's inequality (see, e.g., \cite{Schneider}) 
asserts then that 
\begin{equation} \label{urysohn}
\vrad(K) \leq w(K) .
\end{equation}
A companion inequality, which is even more elementary, is
\begin{equation} \label{dualurysohn}
\vrad(K) \geq w(K^\circ)^{-1} .
\end{equation}
The proof of (\ref{dualurysohn}) is based on expressing the volume as 
an integral in polar coordinates and then using twice H\"older inequality:
$\vrad(K) = 
\left( \int_{S^{m-1}} \|x\|_{K}^{-m} dx \right)^{1/m} 
\geq \int_{S^{m-1}} \|x\|_{K}^{-1} dx 
\geq  \left( \int_{S^{m-1}} \|x\|_{K} dx \right)^{-1} 
= w(K^\circ)^{-1} .
$

\medskip
Applying (\ref{urysohn})  in our setting of the $N^4-1$-dimensional space
$H^{\rm b}$
and for $K= \mathcal{D}_N^{\rm b}$, we obtain
\begin{equation} \label{vradd}
\vrad(\mathcal{D}_N^{\rm b}) \leq  w(\mathcal{D}_N^{\rm b}) 
=w(\co (\mathcal{CP}_N^{\rm b}  \cup \mathcal{C}c \mathcal{P}_N^{\rm b}) )
\leq 
w(\mathcal{CP}_N^{\rm b}  + \mathcal{C}c \mathcal{P}_N^{\rm b}) = 
2 w(\mathcal{CP}_N^{\rm b}) \leq 4 ,
\end{equation}
because $w(\cdot)$ commutes with the Minkowski addition (of sets),
and because $w(\mathcal{CP}_N^{\rm b}) = w(\mathcal{C}c \mathcal{P}_N^{\rm
b})\leq 2$. 
The latter is a consequence of similar estimates for the set 
of all states (which is equivalent to $\mathcal{CP}_N^{\rm b}$ up to a
homothety), 
see \cite{Sza04,AS}.  (We note that while the {\em limit relation}
$w(\mathcal{CP}_N^{\rm b}) \raw 2$
as $N \raw \infty$ follows easily from well-known facts about random matrices,
the {\em estimate valid for  all} $N$ requires finer arguments such as those presented 
in appendices of \cite{Sza04}).  Combining the above estimate with part (i) 
of Theorem 3 we obtain the upper estimate in part (v) with the asserted
constant 8.

The same bound $w(\mathcal{D}_N^{\rm b}) \leq 4$  combined with
(\ref{dualurysohn})
(applied this time with $K= \mathcal{T}_N^{\rm b}$) and with part (i) leads to
the lower bound 
$\frac 14$ in part (iv).

To obtain the asymptotic bounds from part (iv) and (v) with 
the required constants $\frac {e^{1/4}}2$ 
and $2 e^{1/4}$ we argue similarly, 
but instead of the universal estimate $w(\mathcal{D}_N^{\rm b}) \leq 4$ 
we use a 
tighter asymptotic bound 
$\limsup_N w(\mathcal{D}_N^{\rm b}) \leq 2$
This bound is a consequence of  classical isoperimetric inequalities
and  the measure concentration phenomenon that they induce
(see, e.g., \cite{MS}): a Lipschitz function on $S^{m-1}$ is 
strongly concentrated around its mean. In particular, if the out-radius of 
$K$ is at most $R$, then 
$\int_{S^{m-1}} \left| \|x\|_{K^\circ} - w(K) \right| dx  \, = \,
O(R/m^{1/2})$.
If $K = \mathcal{CP}_N^{\rm b}$ or $\mathcal{C}c \mathcal{P}_N^{\rm b}$, then,
by Corollary \ref{radii}, $R=(N^2-1)^{1/2}$ while $m=N^4-1$, hence 
$R/m^{1/2} = (N^2+1)^{-1/2} < N^{-1}$. It is then an elementary exercise to
show
that 
\begin{eqnarray*}
w(\mathcal{D}_N^{\rm b}) &= &\int_{S^{m-1}} 
\max \left(\|x\|_{(\mathcal{CP}_N^{\rm b})^\circ}, \|x\|_{(\mathcal{C}c
\mathcal{P}_N^{\rm b})^\circ}\right) dx\\
&\leq &\max \left(w(\mathcal{CP}_N^{\rm b}), w(\mathcal{C}c \mathcal{P}_N^{\rm
b})\right)
+ O(N^{-1}) \\
&\leq & 2+ O(N^{-1}) ,
\end{eqnarray*}
whence $\limsup_N w(\mathcal{D}_N^{\rm b}) \leq 2$, as required.  
Universal (as opposed to asymptotic) upper bounds on $\vrad(\mathcal{D}_N^{\rm
b})$
better than 4 obtained in (\ref {vradd})  can also be derived this way, most
efficiently by 
converting spherical integrals to Gaussian integrals and using the Gaussian 
isoperimetric inequality. (This would also improve 
somewhat the bounds $\frac 14$ and 8 in parts (iv)
and (v), but we will not pursue this direction here as the 
payoff doesn't seem to justify the effort.)

\smallskip 
Finally, to obtain the {\em lower} bound on $\vrad(\mathcal{P}^{\rm b}_N)$ from
part (ii) 
of the Theorem, we note 
that the upper bound $4N^{-1/2}$ for 
$\vrad(\mathcal{SP}_N^{\rm b})$ (stated in part (iii)) 
was {\em de facto}  (see \cite{AS}) 
deduced from  the stronger estimate $w(\mathcal{SP}_N^{\rm b}) \leq 4N^{-1/2}$. 
It then remains to apply (\ref{dualurysohn}) and the duality between 
$\mathcal{P}_N^{\rm b}$ and $\mathcal{SP}_N^{\rm b}$.

         \subsection{Symmetrized bodies and order intervals} \label{appB}

\bigskip
We will now analyze the polar of the symmetrized body $\calc^{\rm sym}$.
Recall the notation of section 3.2: 
$\calv$ is a real Hilbert space, $\calc \subset \calv$ a closed convex cone,  
$\calc^*$ the dual cone. Next,  
 $e \in \calc \cap \calc^*$ is a unit vector, 
$V^{\rm b} := \{x \in \calv \; : \; \bra x,e \ket =1\}$ is an affine subspace
of $\calv$ 
and
$\calc^{\rm b} = \calc \cap V^{\rm b}$ is the base of the cone $\calc$. 
Finally, the symmetrized body is defined as $\calc^{\rm sym} 
:= \co (-{\mathcal {C}}^{\rm b} \cup {\mathcal {C}}^{\rm b} )$; the minus sign
referring 
to  the symmetric image with respect to $0$.  
An important point, following from classical results \cite{RS, RS2}
and explained in Appendix C of \cite{Sza04}, is that under mild assumptions
which are satisfied 
for all the cones we consider, the volume radii of $\mathcal{C}^{\rm b}$ and of
$\calc^{\rm sym}$
differ by a factor smaller than 2.

Our main assertion (equation (\ref{order}) in section 3.2) is that 
$$\left(\calc^{\rm sym}  \right)^\circ 
= ( e -{\mathcal{C}}^* ) \cap (-e+{\mathcal{C}}^*),
$$
where the polarity has the standard meaning (i.e., inside the entire 
space $\calv$ {\em and} with respect to the origin). That is, 
$y \in \left(\calc^{\rm sym}  \right)^\circ$ iff both $y+e$ and $e-y$ are 
in ${\mathcal {C}}^*$ or, in other words, iff $y$ belongs to 
 $[-e, e]$, the order interval in the sense of the 
order induced by the cone ${\mathcal {C}}^*$.  
For example, if we want to investigate $\mathcal{P}^{\rm b}$ and
$\mathcal{P}^{\rm sym}= \co\left( -\mathcal{P}^{\rm b} \cup \mathcal{P}^{\rm
b}\right)$, we may
specify the framework above to $\mathcal{C} = 
\mathcal{P}$,
obtaining
$$
\left(\mathcal{P}^{\rm sym}\right)^\circ
=\big(-\Phi_*+\mathcal{SP}\big) \cap \big(\Phi_* -\mathcal{SP}\big). 
$$

To prove the assertion, denote $V^{-} := \{x \in \calv \; : \; \bra x,e \ket
\leq 1\}$
(one of the half-spaces determined by $V^{\rm b}$) and 
${\mathcal {C}}^{\rm E} = {\mathcal {C}} \cap { {V}}^{-}$ (cf. Figure 2 
in section \ref{prelim}). Then 
$$\calc^{\rm sym} 
= \co (-{\mathcal {C}}^{\rm b} \cup {\mathcal {C}}^{\rm b} )=
 \co \left(-{\mathcal {C}}^{\rm E}
 \cup {\mathcal {C}}^{\rm E} \right).
$$
Hence, using standard rules for polar operations (see, e.g., \cite{Schneider}),
$$
\left(\calc^{\rm sym}  \right)^\circ 
= \big(-{\mathcal {C}}^{\rm E}\big)^\circ \cap \big({\mathcal {C}}^{\rm E}\big)^\circ.
$$
Next,
$$
\big({\mathcal {C}}^{\rm E}\big)^\circ =
\big({\mathcal {C}} \cap {{V}}^{-} \big)^\circ=
\overline{\co}\left(({{V}}^{-})^\circ \cup \, {\mathcal{C}}^\circ \right) 
= \overline{\co}\left((-\infty,1]\cdot e \ \cup \, -{\mathcal{C}}^* \right)
=e -{\mathcal{C}}^*,$$
where the bar stands for the closure. 
Combining this with the preceding formula and again using the standard 
rules gives
$$
\left(\calc^{\rm sym}\right)^{\circ} = ( e -{\mathcal{C}}^* ) \cap
(-e+{\mathcal{C}}^*)
$$
or the intersection of two cones with vertices at $e$ and $-e$. Clearly
this does not equal $({\mathcal{C}}^*)^{\rm sym} $ except in dimension 1.
However,
the two bodies are closely related. For example, if $e$ is the point of
symmetry of ${\mathcal{C}}^{\rm b}$, then  $({\mathcal{C}}^*)^{\rm sym}$ is a
cylinder
with the base  $({\mathcal{C}}^*)^{\rm b}$ and the axis $[-e, e]$, while 
$\left(\calc^{\rm sym}\right)^{\circ} $ is a union of two cones whose
common base is $\left({\mathcal{C}}^*\right)^{\rm b} - e$, the central section
of the
cylinder, and the vertices are $-e$ and $e$. The two bodies only differ
in one dimension; if thought of as unit balls with respect to the
corresponding norms, the two norms coincide on the hyperspace 
${ \mathcal {V}}_0 := \{x \in \calv \; : \; \bra x,e \ket =0\}$ and on the
complementary one-dimensional space $\R e$, but on the entire space we have
in the first case the direct sum in the $\ell_\infty$ sense, while in the
second case in the $\ell_1$ sense. If the base ${\mathcal{C}}^{\rm b}$ is 
non-symmetric, the situation is
more complicated. For example, the section 
${ \mathcal {V}}_0 \cap \left({\mathcal{C}^{\rm sym}} \right)^\circ$ 
is congruent to the intersection of $({\mathcal{C}}^*)^{\rm b}$ with
its symmetric image with respect to $e$,  but (see \cite{spingarn}) 
the volume radii of the two bodies are comparable if, for example, $e$ is the
only point that is
fixed under isometries of $({\mathcal{C}}^*)^{\rm b}$ (as is the case in all
our applications), or just the centroid of $({\mathcal{C}}^*)^{\rm b}$.

\newpage
  \subsection{Proof of Proposition 2: for ``balanced" cones, $\calc^{\rm b}$
and $\calc^{\rm TP}$ have
 comparable volume radius} \label{appC}
We may assume that $a=0$
(otherwise consider $K-a$).  By hypothesis, we have then
\begin{equation} \label{r-ball}
r B_2^m \subset K \subset R B_2^m,
\end{equation}
where $B_2^m$ is the $m$-dimensional unit Euclidean ball. 
For a subspace $E$, denote by
$P_E$ the orthogonal projection onto $E$.
Then (see \cite{spingarn,mp}),
\begin{equation} \label{mp}
\mbox{vol}_m (K) \leq \mbox{vol}_k(K \cap H)
\ \ \mbox{vol}_{s}(P_{H^\perp} K),
\end{equation}
where $s=m-k$ and  $H^\perp$ is the $m-k$-dimensional space orthogonal to the 
$k$-dimensional subspace $H$.
Therefore
$$\frac{\mbox{vol}_m(K)}{\mbox{vol}_m(B_2^m)}
\ \leq
\ \frac{\mbox{vol}_k(K \cap H)}{\mbox{vol}_k(B^k_{2})}
\ \
\frac{\mbox{vol}_{s}(P_{H^\perp} K)}{\mbox{vol}_s(B^s_{2})}
\ \
\frac{\mbox{vol}_k(B^k_{2})\mbox{vol}_s(B^s_{2})}{\mbox{vol}_m(B^m_{2})}
$$
Hence, using (\ref{r-ball}),
$$
{\vrad (K)^m} \ \leq
\ \vrad (K \cap H)^k \ R^s \ 
\frac{\mbox{vol}_k(B^k_{2})\
\mbox{vol}_{s}(B^{s}_{2})}{\mbox{vol}_m(B^m_{2})} ,
$$
which is the first inequality in (\ref{vradsection}).  
For the second inequality, we start 
with the even more classical result (see \cite{RS2} or \cite{chakerian};
same notation as (\ref{mp}))
\begin{equation}  \label{rs1}
\mbox{vol}_m(K) \geq {m \choose k}^{-1} \mbox{vol}_k(K \cap H)
\ \mbox{vol}_{s}(P_{H^\perp} K) ,
\end{equation}
which doesn't even require that $H$ passes through the centroid of $K$.
As above, this can be rewritten in terms of volume radii as
$$
{m \choose k} {\vrad (K)^m} \ \geq
\ \vrad (K \cap H)^k \ r^s \ 
\frac{\mbox{vol}_k(B^k_{2})\
\mbox{vol}_{s}(B^{s}_{2})}{\mbox{vol}_d(B^m_{2})} ,
$$
which is the second inequality in (\ref{vradsection}).

	\subsection{``No duality" for $\mathcal{CP}_N^{\rm TP}$} \label{appD}
	
The purpose of this Appendix is to show that, in contrast to the bases 
$\mathcal{CP}^{\rm b}$,
the sets $\mathcal{CP}^{\rm TP}$ are very far from being self-dual
in the sense of (\ref{dualbase}), that is, that the polar of 
$\mathcal{CP}^{\rm TP}$ inside the space defined by the trace preserving 
condition (\ref{tpcond}) considered as a vector space with $\Phi_*$ 
as the origin is quite different from the reflection of $\mathcal{CP}^{\rm TP}$ 
with respect to $\Phi_*$.

Generally, if $K \subset \mathbb{R}^m$ is a convex body containing 
the origin in its interior and $H \subset \mathbb{R}^m$ is a vector subspace, 
$K^\circ \cap H$ is always contained in the polar of $K \cap H$ 
inside $H$, and  the discrepancy between the two 
(i.e., the smallest constant $\lambda \geq 1$ such that  the polar of  $K \cap
H$
is contained in $\lambda (K^\circ \cap H)$) is the same as the 
discrepancy between  $K \cap H$ and the orthogonal projection of 
$K$ onto $H$. That discrepancy is also equal to the maximal ratio
between 
\begin{equation} \label{nopolarity}
\max_{x \in K} \bra u,x \ket  \ \ \  {\rm and }\ \ \  \max_{y \in K\cap H} \bra
u,y \ket 
\end{equation}
over nonzero vectors $u \in H$.

In our case $K=\mathcal{CP}_N^{\rm b}$ and $K\cap H =\mathcal{CP}_N^{\rm TP}$.
As a vector space, $H$ may be identified 
with maps whose dynamical matrix has partial trace equal to $0$.
We will argue in the language of dynamical (Choi) matrices considered as 
``flat" block matrices.  In these terms, membership 
in $H$ is equivalent to each block being of trace $0$. 
We will choose  as $u$ the block matrix whose $11$-th block is 
$U=E_{11} - N^{-1}{I}_N$
and the remaining blocks are 0.
Further, we will choose as $x$ the matrix whose $11$-th block is 
$X=NE_{11} $ and the remaining blocks are 0; then the scalar product
corresponding to 
$\bra u,x \ket$ is $\tr (UX) = N-1$.
On the other hand,  if $Y$ is the $11$-th block of the Choi matrix 
of any element of $\mathcal{CP}_N^{\rm TP}$, then $Y$ is a state 
and so the scalar product corresponding to 
$\bra u,y \ket$ is $\tr (UY) = \tr (E_{11}Y) - N^{-1}\tr Y \leq \tr Y -
N^{-1}\tr Y = 1-N^{-1}$.
Accordingly, the discrepancy between the two maxima in (\ref{nopolarity})
is at least $(N-1)/(1-N^{-1})=N$.

\subsection{Volume radius of the set of trace non increasing maps}
\label{appE} 

We want to  determine  the asymptotic order of the volume radius of the set of
all completely
postive, trace non increasing maps $\Phi: \mathcal{M}_N \rightarrow
\mathcal{M}_N$ i.e. the set
$$\mathcal{C P}_N^{\rm TNI} := \{ \Phi \in
\mathcal{C P}_N: \ \mbox{Tr}\, \Phi(\rho) \leq \mbox{Tr}
\, \rho \ \ \mbox{for all} \ \ \rho \geq 0\}.$$
As pointed out earlier, an exact formula for that volume was very recently
found
(independently from this work and by a different method) in \cite{CSZ07}.
However, an argument using the approach of this paper 
is conceptually very simple and so we include it.  We have

\begin{prop} \label{prop6}
We have, for all $N$, 
\begin{equation} \label{tni}
\big({e\, N^{5/2}}\big)^{-N^2}  \leq
\frac{\vol \big(\mathcal{C P}_N^{\rm TNI}\big)}{\vol \big(\mathcal{C P}_N^{\rm
TP}\big)
\times \vol \big( \{M \in
\mathcal{M}_N: 0 \leq M \leq {I}_N\}\big) } \leq N^{-N^2/2}
\end{equation}
\end{prop}
To derive estimates on  $\vol \big(\mathcal{C P}_N^{\rm TNI}\big)$ from the
Proposition, 
one needs to use the readily available information on the two factors in the 
denominator of the middle term of (\ref{tni}). First, the asymptotic order of
the 
volume radius of $\mathcal{CP}_N^{\rm TP}$ was determined in Theorem
\ref{thm5}(i).
Next, the set   $\mathcal{A} := \{M \in
\mathcal{M}_N: 0 \leq M \leq {I}_N\}$ is a ball  of radius $1/2$ (in
the operator norm)
centered at ${I}_N/2$, and so its volume radius admits easy bounds 
given by the in- and outradius: $1/2$ and $\sqrt{N}/2$ (actually a much tighter
lower bound 
$\sqrt{N}/4$ can be obtained via a slight modification of the argument from
Theorem \ref{thm2}(i),  
see Appendix \ref{appA}, but for our purposes the trivial bounds suffice).
A straighforward calculation leads then to
\begin{cor} \label{cor7}
$$
\lim_{N \raw \infty} \vrad\big(\mathcal{CP}_N^{\rm TNI}\big) = e^{-1/4} .
$$
\end{cor}
The key point is that  in order to calculate the volume radius we need to 
raise the volume to the power $1/N^4$. Thus
the factors such as  $(e\, N^{5/2})^{-N^2}$ on the left hand side of (\ref{tni})
are
inconsequential since
it leads to an expression of the form 
$1- O(\frac{\log{N}}{N^2})$.
For the same reason, the effects of $ \vol (\mathcal{A}) $ and of the 
$b(m,k)$-type factor, which also enters the calculation 
(cf. Proposition \ref{section} and the comments following it),  
tend to 0 as $N \raw \infty$.

\medskip
For the proof of Proposition \ref{prop6}  we note first that
$\mathcal{CP}_N^{\rm TNI}$
is canonically isometric to  the set of {\em subunital} maps
$$\mathcal{C P}_N^{\rm SU} := \{ \Phi \in \mathcal{C P}: \ \Phi({I}_N)
\leq {I}_N\}.$$
In what follows we will work with the latter set. 
The isometry, which assigns to $\Phi : \calm_N \raw \calm_N$ the dual (in the
linear algebra, or Banach 
space sense) map $\Phi^*$, sends $\mathcal{CP}_N^{\rm TP}$ to the set of  
{\em unital} maps 
$\mathcal{C P}_N^{\rm U} := \{ \Phi \in \mathcal{C P}_N: \ \Phi({I}_N)
= {I}_N\}.$
The set $\mathcal{C P}_N^{\rm SU}$ admits a natural fibration: with every 
$M \in \mathcal{A}$, we may associate 
\begin{equation} \label{F_M}
F_M=\{\Phi \in \mathcal{C P}: \ \Phi({I}_N) =M\};
\end{equation}
in particular 
$F_{{I}_N}=\mathcal{C P}_N^{\rm U}$.
In the language of Choi (dynamical) matrices  the equality from (\ref{F_M}) 
translates to $\tr_B D_\Phi = M$, or  to $\sum_j D_{jj} =M$ if we think of
$D_\Phi$ 
as a block matrix  $D_\Phi=(D_{jk}) = \big(\Phi(E_{jk})\big)$. Since all fibers
$F_M$ are parallel
to the subspace $\mathcal{N}$ defined by $\tr_B D_\Phi = 0$ (or $\sum_j D_{jj}
=0$),
one can express the volume as an integral
\begin{equation} \label{fubini}
\vol \big(\mathcal{C P}_N^{\rm TNI}\big)= \vol \big(\mathcal{C P}_N^{\rm
SU}\big)= 
N^{-N^2/2}
\int_{\mathcal{A}} \mbox{vol}\big(F_M\big) \, dM 
\end{equation}
The reason for the factor $N^{-N^2/2}$ is that while the fibration is naturally 
parametrized by the elements of $\mathcal{A}$, the projection of $F_M$ 
onto $\mathcal{N}^\perp$ is actually the map $\rho \raw N^{-1} \tr (\rho) M$,
whose Choi 
matrix is $N^{-1} {I}_N \otimes M$ (or a block matrix whose all diagonal
blocks are 
$M/N$ and off-diagonal blocks are 0). Now, the Hilbert-Schmidt norm of $N^{-1}
{I}_N$ is 
$N^{-1/2}$, and so the projection of $\mathcal{C P}_N^{\rm SU}$ onto
$\mathcal{N}^\perp$ 
is isometric to $N^{-1/2}\mathcal{A}$.

The second inequality in (\ref{tni}) is now an immediate consequence of 
(\ref{fubini}) and the bound 
$\vol\big(F_M\big) \leq \vol\big(F_{{I}_N}\big)=\vol\big(\mathcal{C
P}_N^{\rm U}\big)=\vol\big(\mathcal{C
P}_N^{\rm TP}\big)$,
valid for all $M \in \mathcal{A}$, which in turn follows, e.g., from $F_M$
being the image of 
$F_{{I}_N}$ under the contraction 
$g_M : \Phi(\cdot) \raw M^{\frac{1}{2}}  \Phi(\cdot)M^{\frac{1}{2}}$. (On the
level of dynamical 
matrices, the action of $g_M$ is given by 
$D_\Phi \raw \big({I}_N\otimes M^{\frac{1}{2}}\big) D_\Phi
\big({I}_N\otimes M^{\frac{1}{2}}\big)$ or, in the language of block
matrices, by 
$(D_{jk}) \raw (M^{\frac{1}{2}}D_{jk}M^{\frac{1}{2}})$.) 
The fact that $g_M\big(F_{{I}_N}\big) \subset F_M$ is obvious from the
definition;
surjectivity for invertible $M$'s follows by considering the inverse
$g_M^{-1} = g_{M^{-1}}$, and for 
singular $M$'s by looking at invertible approximants.

\vskip 3mm
For the first inequality in (\ref{tni}) we may use (\ref{rs1}) with
$K=\mathcal{C P}_N^{\rm SU}$ and
the section $K\cap H = \mathcal{C P}_N^{\rm U}$. As pointed out earlier, the
set  $P_H K 
= P_{\mathcal{N}^\perp}$ is then isometric to $N^{-1/2}\mathcal{A}$, and it
remains to use the elementary 
bound ${m \choose k}={m \choose s} \leq (em/s)^s$. A alternative argument is to restrict the
integration 
in (\ref{fubini}) to $\{M : t {I}_N \leq M \leq {I}_N\}$, which
is a ball in the operator norm of radius $(1-t)/2$, then use the fact that for
such 
$M$ the function $c g_M^{-1}$ is a contraction, and finally optimize over $t
\in (0,1)$.
This approach allows in fact to express the Jacobian of $g_M$ in terms of
eigenvalues of 
$M$ and, subsequently, to express the ratio under consideration 
as a multiple integral over $[0,1]^N$,
but we will not pursue this path further.

%%%%%%%%%%%%%%%%%%%%%%%%%%%%%%%%

\end{document}